\documentclass[useAMS,usenatbib]{mn2e}
\usepackage{epsfig,subfigure}
\usepackage{amssymb}

\newcommand{\rperi}{\ensuremath{r_{\rm peri}}}

\begin{document}

\title[Dark Subhaloes and Extended HI Discs]{Dark Subhaloes and Disturbances in Extended HI Discs}

\author[Chang \& Chakrabarti]{Philip Chang$^1$\thanks{E-mail:
pchang@cita.utoronto.ca (PC); schakra1@fau.edu (SC)} \& Sukanya
Chakrabarti$^{2}$
\\$^1$Canadian Institute for Theoretical Astrophysics, 60 St George St, Toronto,
ON M5S 3H8, Canada
\\$^2$Department of Physics, Florida Atlantic University, 777 Glades Road, Boca Raton FL 33431-0991} 

\maketitle

\begin{abstract}
We develop a perturbative approach to study the excitation of disturbances in the extended 
atomic hydrogen (HI) discs of galaxies produced by passing dark matter subhaloes.  The shallow gravitational potential of the dark matter subhaloes (compared to the primary halo) allows us to use the epicyclic approximation, 
the equations of which we solve by modal analysis, i.e., assuming a disc is composed of N radial rings with M modes.  
We show that properties of dark matter subhaloes can be inferred from the profile and amplitude of the modal energy of the disc.
Namely, we find that the overall amplitude of the response gives the mass of the dark sub-halo. Motivated by this modal analysis, we then show that the density response shows similar features. Finally, we show that our results agree with those from full hydrodynamic simulation.  
We find a simple scaling relation between the satellite mass and Fourier amplitudes of the resultant surface density of the gas disc where the effective Fourier amplitude (essentially a sum over the low order modes) scales as $m_{s}^{1/2}$, where $m_{s}$ is the satellite mass.  The utility of this relation is that it can be readily applied to an observed HI map to deduce the satellite mass without recourse to full numerical simulation.  This will greatly aid us in analyzing large samples of spiral galaxies to constrain the population of dwarf satellites in the Local Volume.

\end{abstract}

\begin{keywords}
{
galaxies: interactions -- dark matter -- galaxies: general
}
\end{keywords}

\section{Introduction}\label{sec:intro}

Dark matter halos grow by a sequence of mergers and accretion events in cold dark matter (CDM) cosmology.  Due to the absence of scale in this heirarchical structure formation scenario, dark matter haloes are expected to be merely scaled-up or down versions of each other with the total mass being the scaling parameter \citep{Navarro+97}.  Observation of galaxies at different masses show that this is clearly not the case.  At the high end, clusters have thousands of galaxies, i.e., subhaloes, in line with theoretical predictions \citep{Natarajan+04}.  However, galactic size halos have a few bright satellites, e.g., our Milky Way (MW), in constrast with theoretical models of the MW halo (e.g. \citealt{Diemand+08}), and dwarfs have no bright satellites.  This qualitative difference at the faint end has come to be known as the substructure or ``missing satellites`` problem \citep{Klypin+99,Kravtsov04,Madau08,Kravtsov10}\footnote{To be precise, the ''missing satellites`` problem refers to the lack of observed satellites in the MW halo compared to what is expected theoretically.}.  

\citet{Kravtsov10} recently review the possible solutions to the ''missing satellites`` problem.  To summarized, solving the ''missing satellites`` problem demands that a scale be introduced either in structure formation of dark matter haloes or in the physics of galaxy formation to suppress the number of observed systems at small scales.  Namely, the power spectrum of fluctuations must be suppressed at small scales via warm dark matter \citep{Zentner+03,Hooper+07,Primack09} or star formation is suppressed in dwarfs due to reionization \citep{Bullock+00,Okamoto+09,Busha+10,Iliev+10} or feedback from the first stars \citep{Wise+08}.

Recent discoveries of dwarf galaxies in the MW and M31 have
relieve some of this tension (for a review see \citealt{Willman10}). These dwarfs are the most dark matter dominated \citep{Strigari+08} and metal poor systems known \citep{Kirby+08}.  While their unusual properties point to the difficulty of forming stars in these systems, the number density of these dwarfs may alleviate the "missing satellite" problem once completeness corrections are accounted for \citep{Tollerud+08}. 

The evidence from the MW observations and theoretical predictions point to the existence of many nearly dark subhaloes hosting few stars.  For instance, \citet{Bullock+10} recently pointed out that these stealth galaxies would be missed in current survey, having surface brightnesses that are too diffuse.  
This motivates finding alternative means to constrain this dark subhalo population.

In a recent series of papers \citep[hereafter CB09 and CB10, respectively]{Chakrabarti+09,Chakrabarti+10}, 
CB09 and CB10 made the assumption that the observed disturbances in the outer atomic hydrogen gas disc of the MW \citep{Levine+06} are due to the tidal interaction with a dark matter dominated dwarf galaxy.  Proceeding under this hypothesis, CB09 and CB10 numerically simulated a suite of encounters with dwarf galaxies interacting with the MW, and developed an approach to infer the mass and current location of satellites from analysis of their tidal imprints on outer gas discs. 
The Fourier amplitudes (CB09) and phase (CB10) of the suite of simulations were 
compared to the observed amplitudes and phases to
determine the pericenter approach, current position (in radius and azimuth), and mass of the subhalo
that drove the observed disturbances.  In doing so, CB09 and CB10 demonstrated how to characterize a dark subhalo from looking at it's effects on the primary galaxy's extended HI disc.  Following this work, we demonstrated the 
validity of this method by performing the analysis with respect to galaxies with $\it{known}$ optical companions \citep[hereafter CBCB]{Chakrabarti+11}.  We showed that the method accurately recovers the mass of the satellite and its position purely from analysis of observed disturbances in the HI disc, 
without requiring any knowledge of the optical light from the satellites.  

In this paper, we continue this work and develop a simplified test particle approach to studying the generation of these disturbances in extended HI discs by dark matter subhaloes.  Using this approach, we develop scaling relations between the satellite mass and Fourier amplitudes of the resultant surface density of the HI disc.  These relations can be utilized by observers who wish to determine the satellite mass directly from the observed HI map, without having to take recourse to full numerical simulations. Our simplified approach is motivated by the fact that DM subhaloes have a much lower velocity dispersion, i.e., gravitational potential, than the host halo in which they sit.  Thus, we calculate the motions of the gas (or stars) in the epicyclic approximation and show that this approximation matches the results of both test body calculations and full numerical simulations.  

This work is complementary to the work of CB09, CB10, and CBCB.  Rather than compare the results of numerical simulations to observations of specific extended HI discs, this work focuses on the physics of the subhalo excitation and exploration of its parameter space.  In the future, we will apply these scaling relationships to a large sample of local spiral galaxies to determine the statistical viablity of the Tidal Analysis method, i.e., the incidence of false positives.  Moreover, we will attempt to constrain the population of dwarf galaxies in the Local Volume.

This paper is organized as follows.  In \S \ref{sec:equations}, we take the equations of motion for a test particle and develop its modal equivalent in the epicyclic approximation.  We then show how the modal viewpoint can be related to the test particle viewpoint.  This modal analysis is shown to reproduce test particle calculations accurately.  In addition, the invariance of the modal energy response (in the absence of dissipation) provides us an unique vantage point to study the response of the disc.  In \S\ref{sec:influence}, we study the modal energy response and the corresponding density response.  We also compute global quantities from these responses as a measure of the global properties of the disturbances raised by the passing DM subhalo in \S\ref{sec:scaling}.  We demonstrate a scaling relation between the disc response and the mass of the satellite where the amplitude of the modes scale like $M_{\rm s}^{1/2}$.  To show that our analysis accurately models the relevant physics, we then compare our results to those computed from large scale simulations in \S\ref{sec:sph}.  The good agreement between the SPH simulations and our results is encouraging.  In addition, the scaling which we derived from our simplified calculation continues to hold.  Finally, we close in \S\ref{sec:conclusions} with a discussion of how these methods can be used to constrain the properties of DM subhaloes in galaxy haloes.

\section{Basic Equations}\label{sec:equations}

We begin with the equations of motion of a test particle in 2-D
($r,\theta$ coordinates):
\begin{eqnarray}
\frac{dr}{dt} &=& v_r \label{eq:1}\\
\frac{d\phi}{dt} &=& \frac{J}{r^2}\\
\frac{\partial v_r}{\partial t} &+&
v_r\frac{\partial v_r}{\partial r} + \frac{J}{r^2}\frac{\partial
  v_r}{\partial \theta} = \frac{J^2}{r^3}-\frac{\partial \Phi}{\partial r}\\
\frac{\partial J}{\partial t} &+&
 v_r\frac{\partial J}{\partial r} + \frac{J}{r^2}\frac{\partial
  J}{\partial \theta} = -\frac {\partial \Phi}{\partial \theta}\label{eq:4},
\end{eqnarray}
where $v_r$ is the radial velocity, $J=\dot{\phi}r^2$ is the angular
momentum, $\phi(t)$ is the $\theta$ position of the particle, and
$\Phi$ is the gravitational potential.  We assume a background
state of circular orbits, i.e., $v_{r,0} = 0$, about
an axissymmetric potential, $\Phi_{0}$.  Expanding equations
(\ref{eq:1}) - (\ref{eq:4}) to linear order, e.g., $r \rightarrow r + \delta
r$, we find
\begin{eqnarray}
\frac{d\delta r}{dt} &=& \delta v_r \label{eq:linear1}\\
\frac{d\delta\phi}{dt} &=& \frac{\delta J}{r^2} - \frac{2 J}{r^2}\frac{\delta
r}{r}\\
\frac{\partial \delta v_r}{\partial t} &+&
\frac{J}{r^2}\frac{\partial
  \delta v_r}{\partial \theta} = \frac{2J\delta J}{r^3} -
\frac{3J^2}{r^3}\frac{\delta r}{r} - \frac{\partial \delta\Phi}{\partial r}\\
\frac{\partial \delta J}{\partial t} &+&
 \delta v_r\frac{\partial J}{\partial r} + \frac{J}{r^2}\frac{\partial
 \delta J}{\partial \theta} = -\frac {\partial \delta \Phi}{\partial
\theta}\label{eq:linear4}.
\end{eqnarray}
We assume that the perturbation to the potential, $
\delta\Phi$, arises from external perturbers ignoring the self-gravity of the particles.  
This simplification is safe for typical HI disc parameters.

Assuming solutions of the form $\exp(im(\theta - \Omega(r)t))$, where
$\Omega(r) = \dot\phi(r)$, e.g.
\begin{displaymath}
\delta r \propto \sum_{m=1}^M\delta r_m(r,t)\exp(im(\theta - \Omega(r)t)),
\end{displaymath}
reduces equations (\ref{eq:linear1}) - (\ref{eq:linear4}) to M
separate wave equations, sourced by the Fourier components of the
perturbing potential:
\begin{eqnarray}
\frac{\partial\delta r}{\partial t} &=& \delta v_r \label{eq:modal1}\\
\frac{\partial\delta\phi}{\partial t} &=& \frac{\delta J}{r^2} - \frac{2
J}{r^2}\frac{\delta r}{r}\\
\frac{\partial \delta v_r}{\partial t} &=& \frac{2J\delta J}{r^3} -
\frac{3J^2}{r^3}\frac{\delta r}{r} - \frac{\partial \delta\Phi}{\partial
r}\exp\left(im\Omega(r)t\right)\\
\frac{\partial \delta J}{\partial t} &=& -\delta v_r\frac{\partial J}{\partial
r} - im\delta \Phi\exp\left(im\Omega(r)t\right).\label{eq:modal4}
\end{eqnarray}
Solving these perturbed equations for a finite $M$ (say $M=10$) allows
us to estimate the linear response of a circular ring of of orbiting
particles to external disturbances, i.e., passing DM subhaloes.  Thus,
any individual particle's perturbed orbit, whose guiding center is at
radius $r$, can be found by summing over this response.  Namely, we reconstruct the particle positions from these modes using:
\begin{eqnarray}\label{eq:reconstruction 1}
r_i(t) = r_{i,0} + \delta r_i(t), \\
\phi_i(t) = \phi_{i,0} + \delta \phi_i(t),\label{eq:reconstruction 2}
\end{eqnarray}
where $r_{i,0}$ and $\phi_{i,0}$ are the initial radius and azimult of the particle, $i$, and 
\begin{eqnarray}
\delta r_i(t) = \sum_{m=1}^M\delta r_m(r_{i,0},t)\exp(im(\phi_{i,0} - \Omega(r_{i,0})t)),\\
\delta \phi_i(t) = \sum_{m=1}^M\delta \phi_m(r_{i,0},t)\exp(im(\phi_{i,0} - \Omega(r_{i,0})t)),
\end{eqnarray}
This greatly simplifies this orbit into a sum of $M$ simple harmonic oscillators,
whose natural frequency only depends on their radial position. 

\begin{figure*}
  \begin{center}
  \subfigure[]{
  \includegraphics[width=0.45\textwidth]{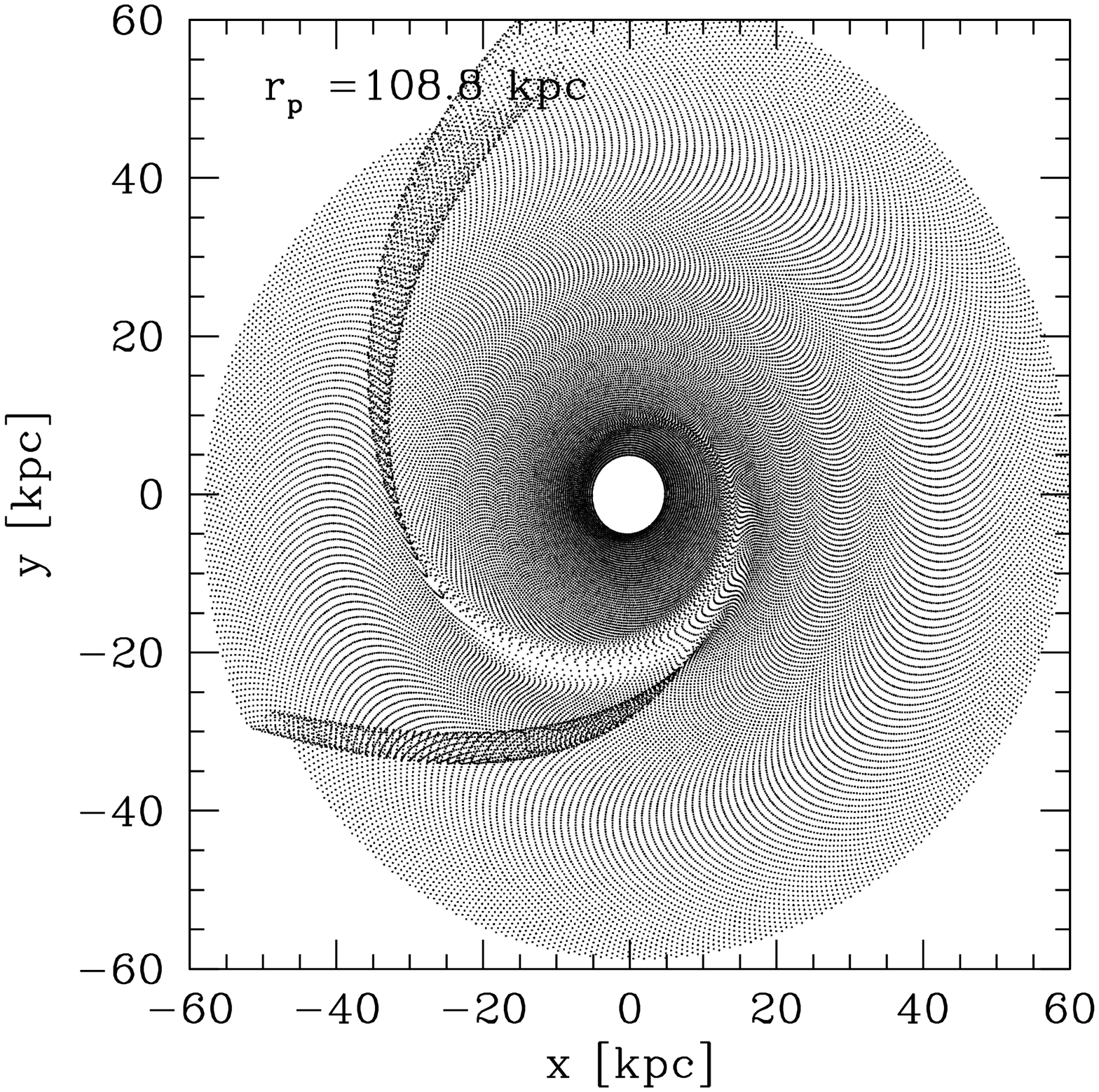}}
  \subfigure[]{
  \includegraphics[width=0.45\textwidth]{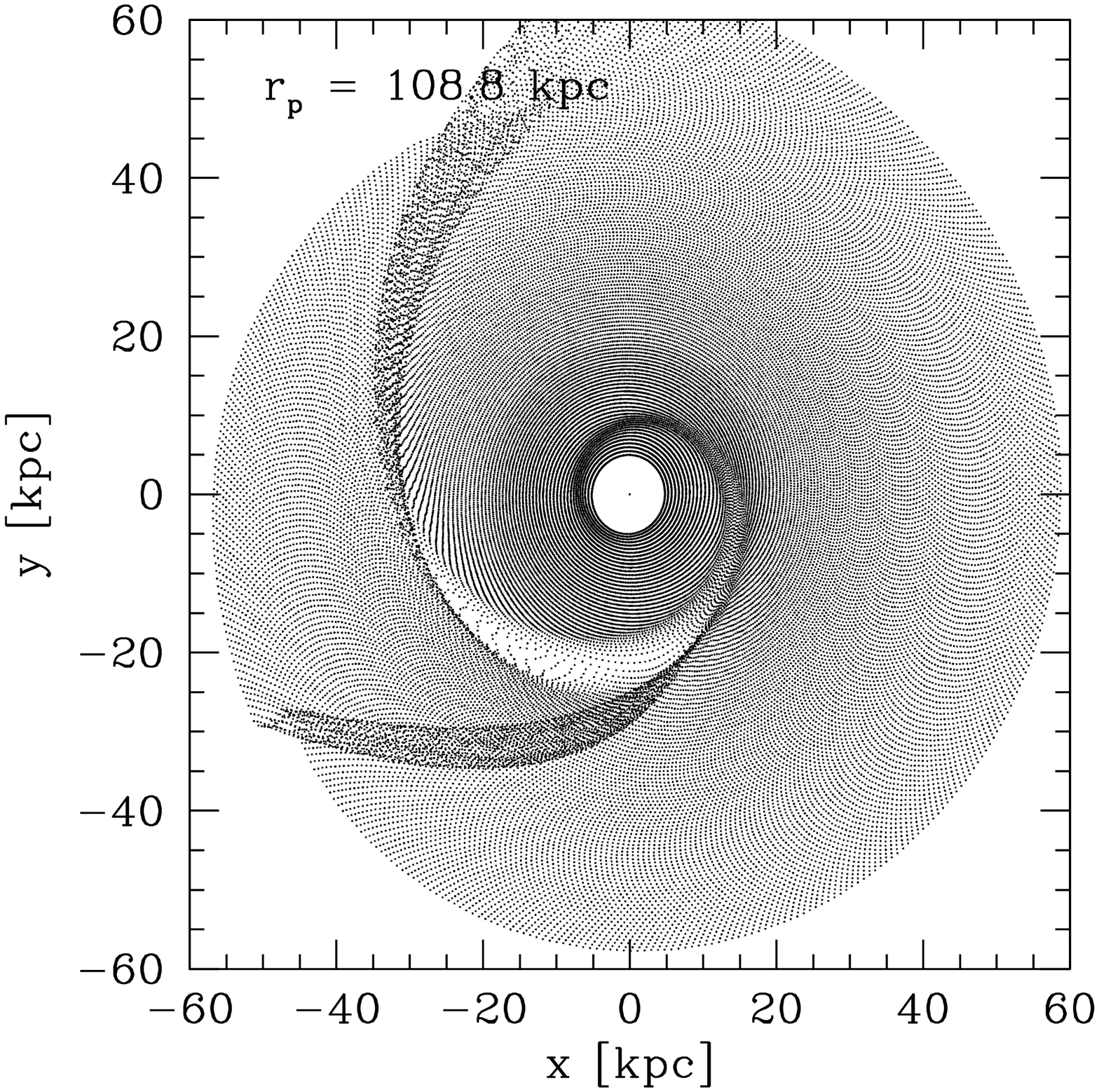}}
  \end{center}
\caption{Comparison between a direct integration (left) and the mode
  calculation (right) of a collisionless particle disc excited by a
  passing DM subhalo.  In this calculation, we use 50000 particles for the test-particle calculation. For the modal calculation, we set $M=64$ for 200 rings.  From this plot it is clear that modal and N-body calculation agree.}
\label{fig:nbody_vs_analytic}
\end{figure*}

To illustrate the power of this formalism, we compare a direct numerical calculation of a collisionless particle disc with that of the modal calculation above and show the results of the direct test particle calculation and our modal reconstruction in Figure
\ref{fig:nbody_vs_analytic}.  Here, the broad features of the response is captured between
the mode calculation and the N-body calculation.  It is evident that the modal calculation captures the details of the perturbation accurately.


Our formalism makes the analogy between the different modes of a radial
ring and a collection of uncoupled simple harmonic oscillators, whose oscillator
frequency just depends on their radius. Hence it is natural to think of the
modal energy, 
\begin{eqnarray}\label{eq:energy}
E_m = \frac 1 2 \delta v_{r,m}^2 + \frac 1 2 \kappa^2 \delta r_m'^2,
\end{eqnarray}
where $\kappa = \sqrt{R(\partial\Omega^2/\partial r + 4\Omega^2}$ is the epicyclic frequency and 
\begin{equation} 
\delta r_m' = \delta r_m - \frac {2 \Omega \delta J_m} {r\kappa^2} 
\end{equation}
is the perturbed radial mode $m$ that accounts for the perturbation to the guiding center.
Looking at the oscillators energy has the major advantage that it remains constant in the absence of dissipation or excitation, whereas the density response of a disc varies with time.  This is advantageous in elucidating the physics.  In the next section, we show the benefit of this mode of analysis and relate its properties to the density response.

\section{Disc Response}\label{sec:influence}

For the purposes of this work, we make a series of simplifying assumptions. First, we assume a static potential for both the primary galaxy and the subhalo, i.e., we do not use a "live halo".  Second, we assume a constant density HI disc motivated by observations of the Galactic HI disc \citet{Wong+02} and the extended HI discs of external galaxies \citet{Bigiel+10}.  There also exists an exponential gas disc, but we ignore this for the sake of simplicity.\footnote{CB09, CB10, and CBCB use an exponential gas disc with an added flat HI disc.}  These assumptions allow us to develop a simplified approach to the modeling of disturbances in extended HI discs.  We demonstrate the accuracy of the results of the simplified approach to that of a full simulation later in \S\ref{sec:sph}.

\subsection{Modal Energy Response}\label{sec:modal energy response}

\begin{figure*}
\begin{center}
\subfigure[]{
\includegraphics[width=.27\textwidth]{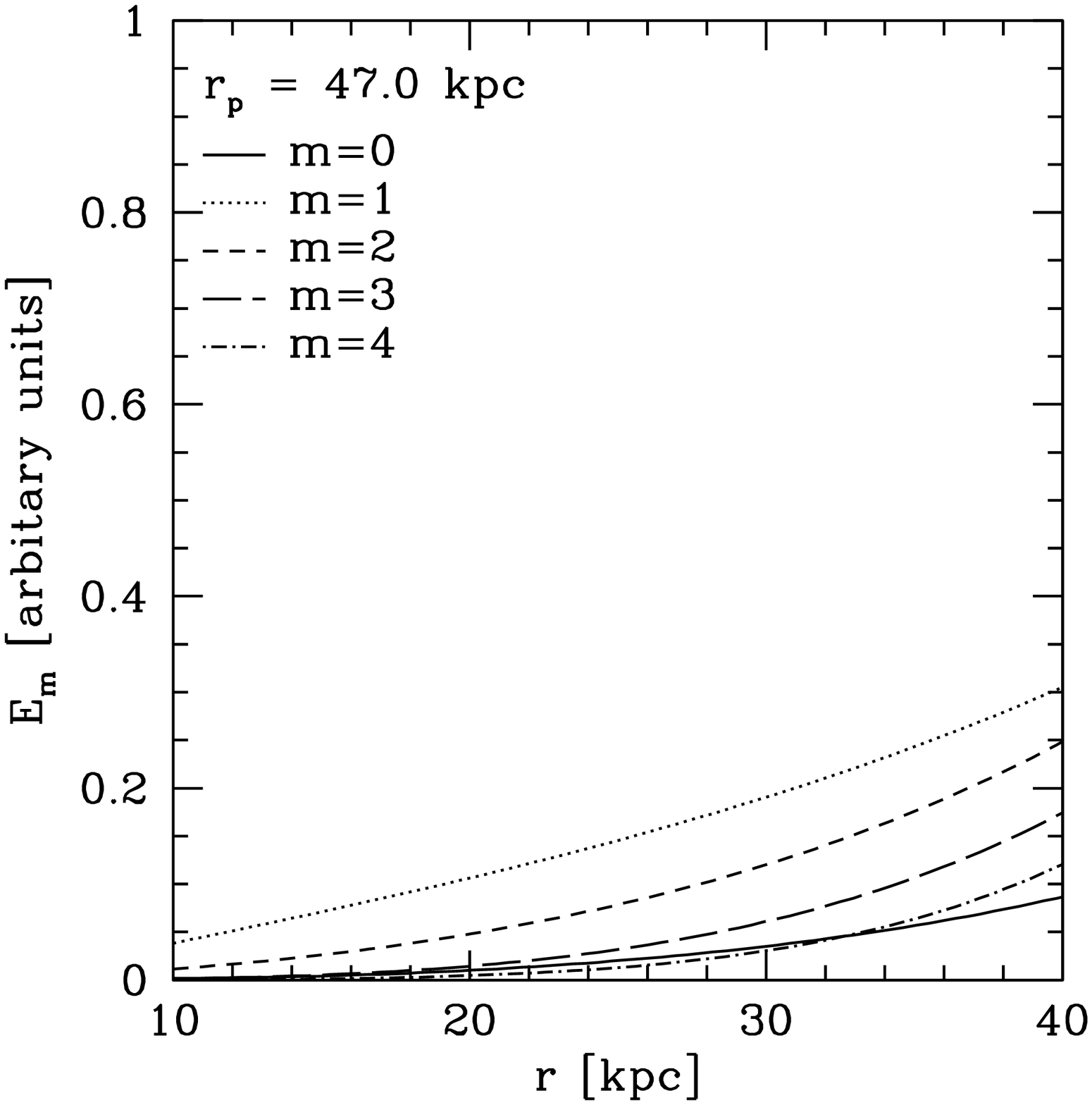}}
\subfigure[]{
\includegraphics[width=.27\textwidth]{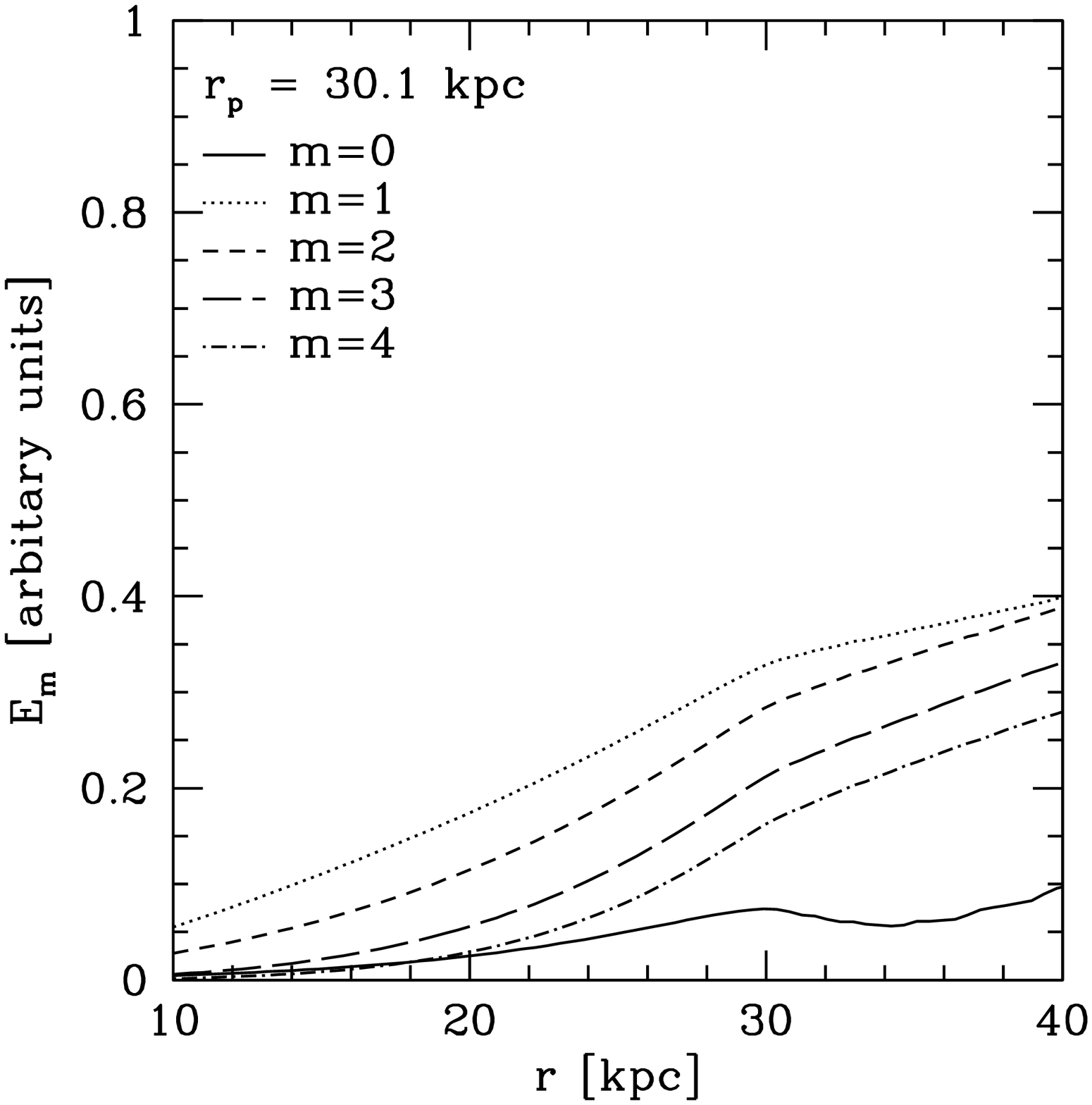}}
\subfigure[]{
\includegraphics[width=.27\textwidth]{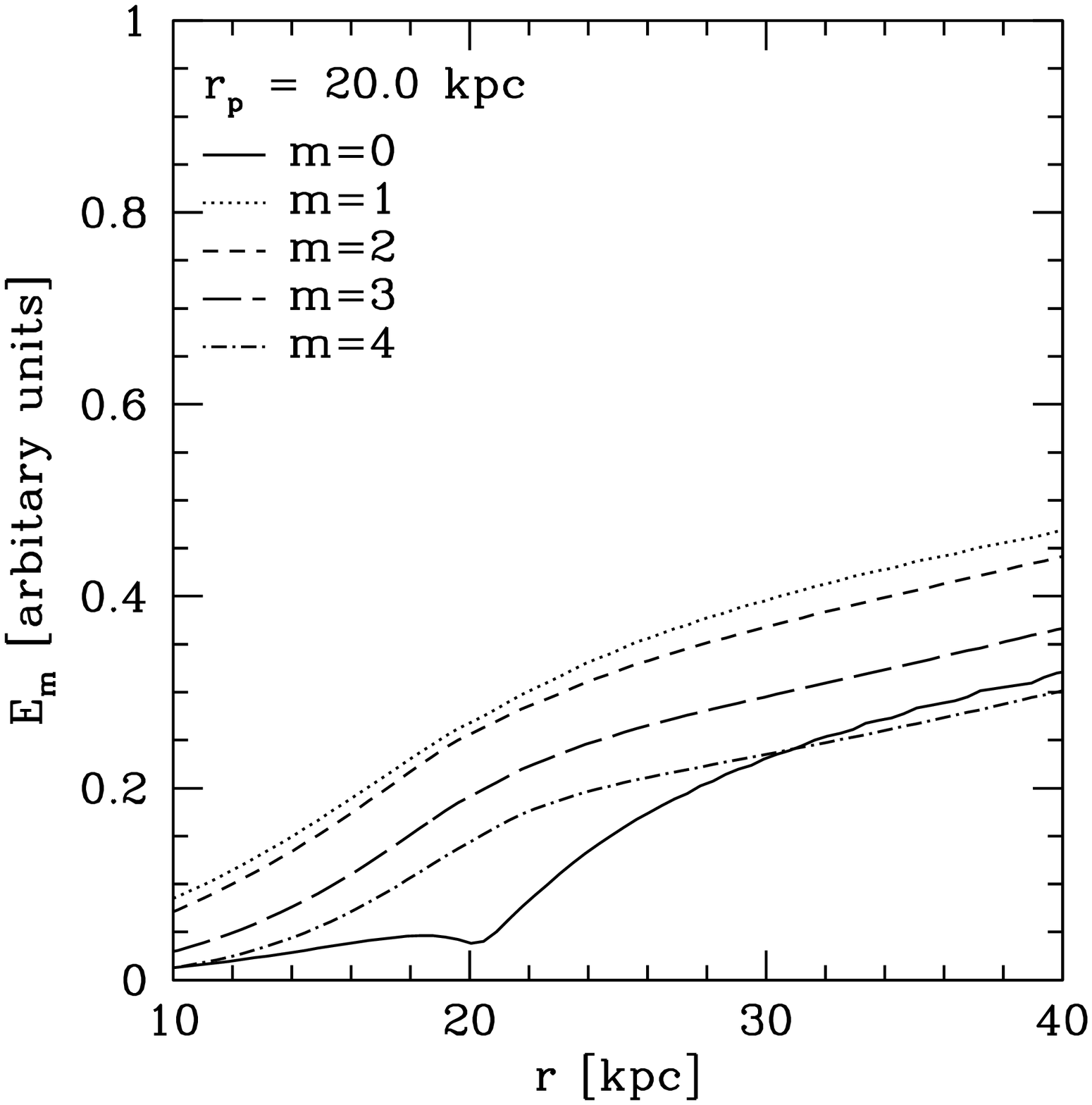}}
\subfigure[]{
\includegraphics[width=.27\textwidth]{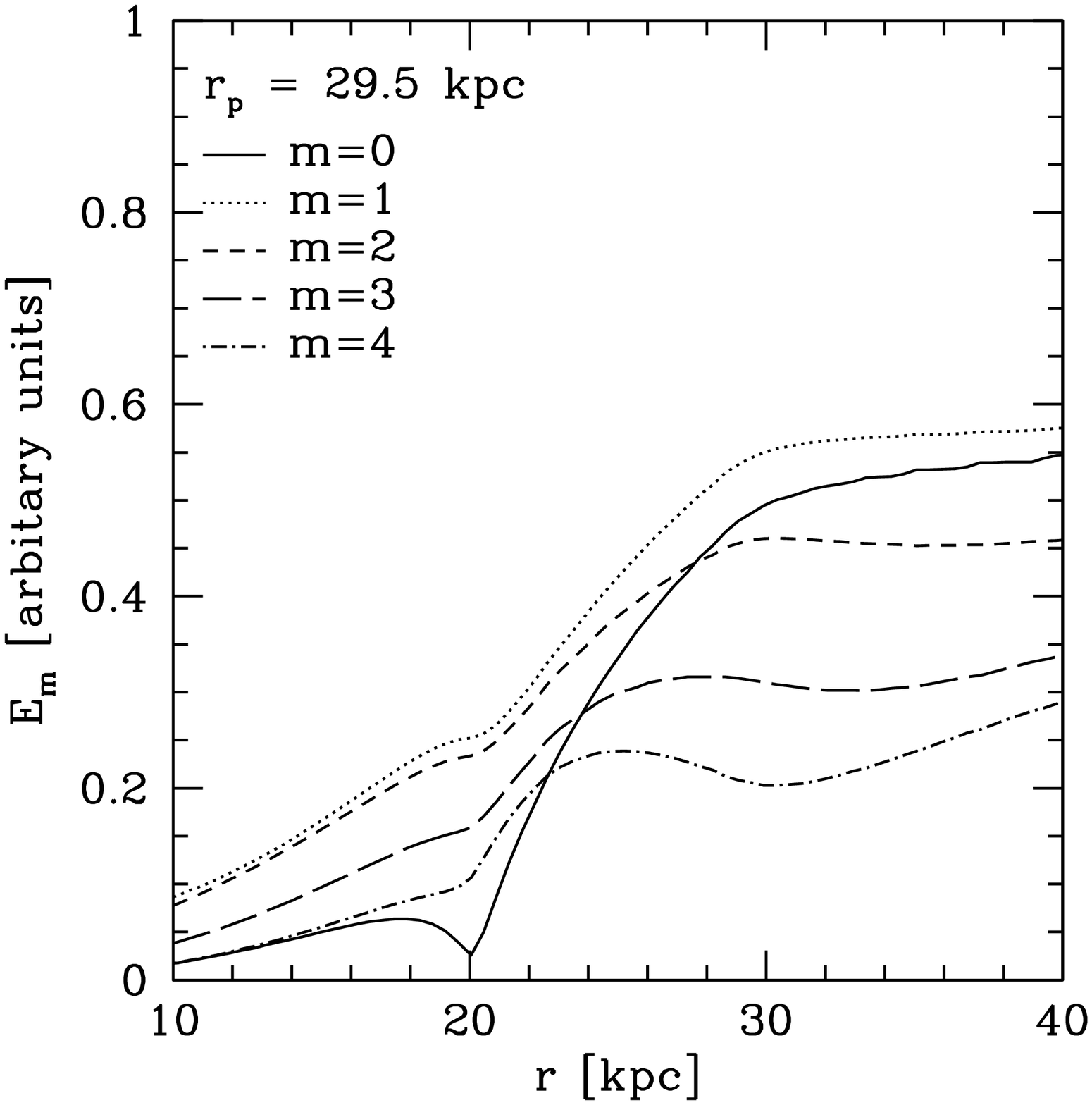}}
\subfigure[]{
\includegraphics[width=.27\textwidth]{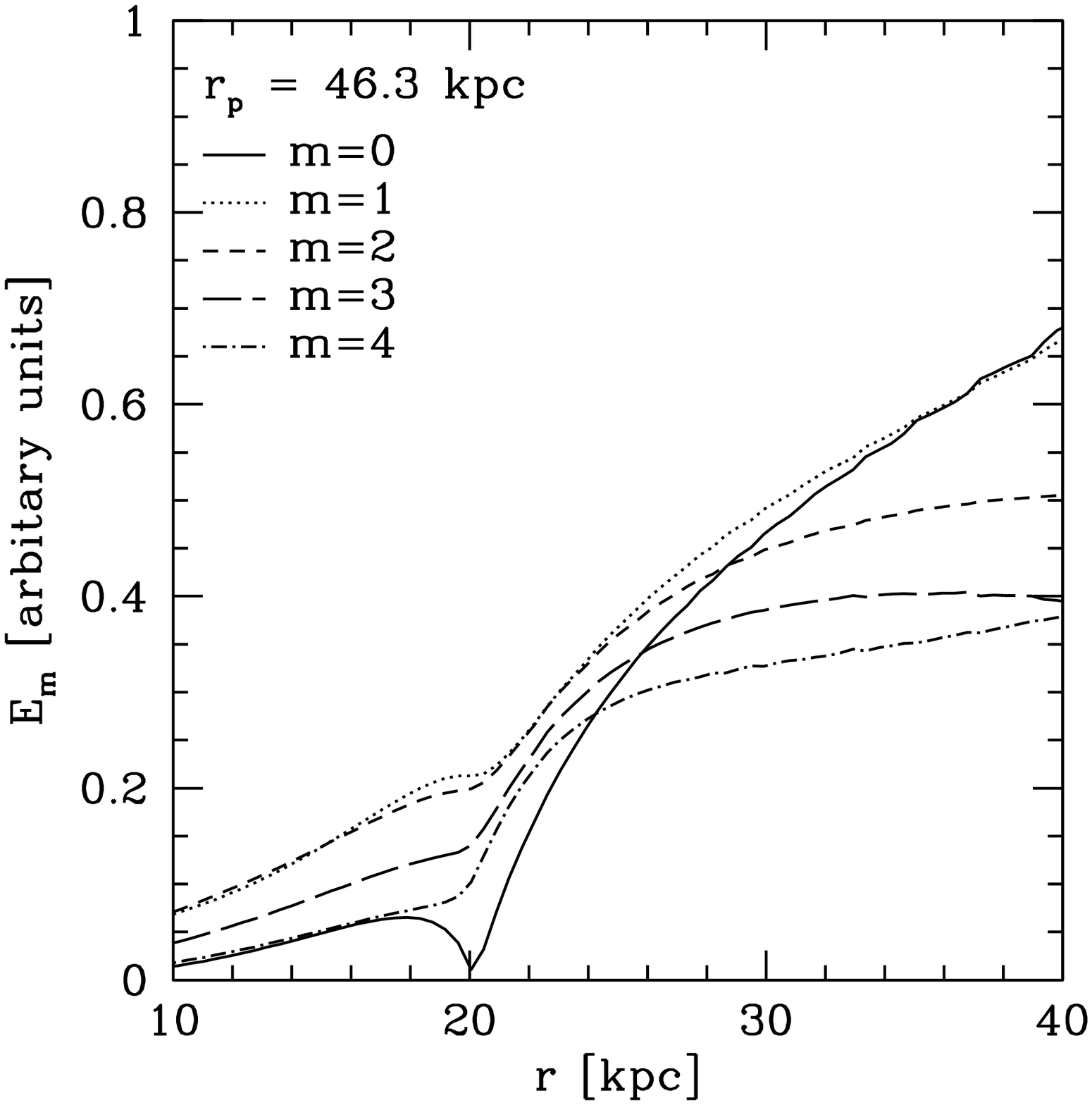}}
\subfigure[]{
\includegraphics[width=.27\textwidth]{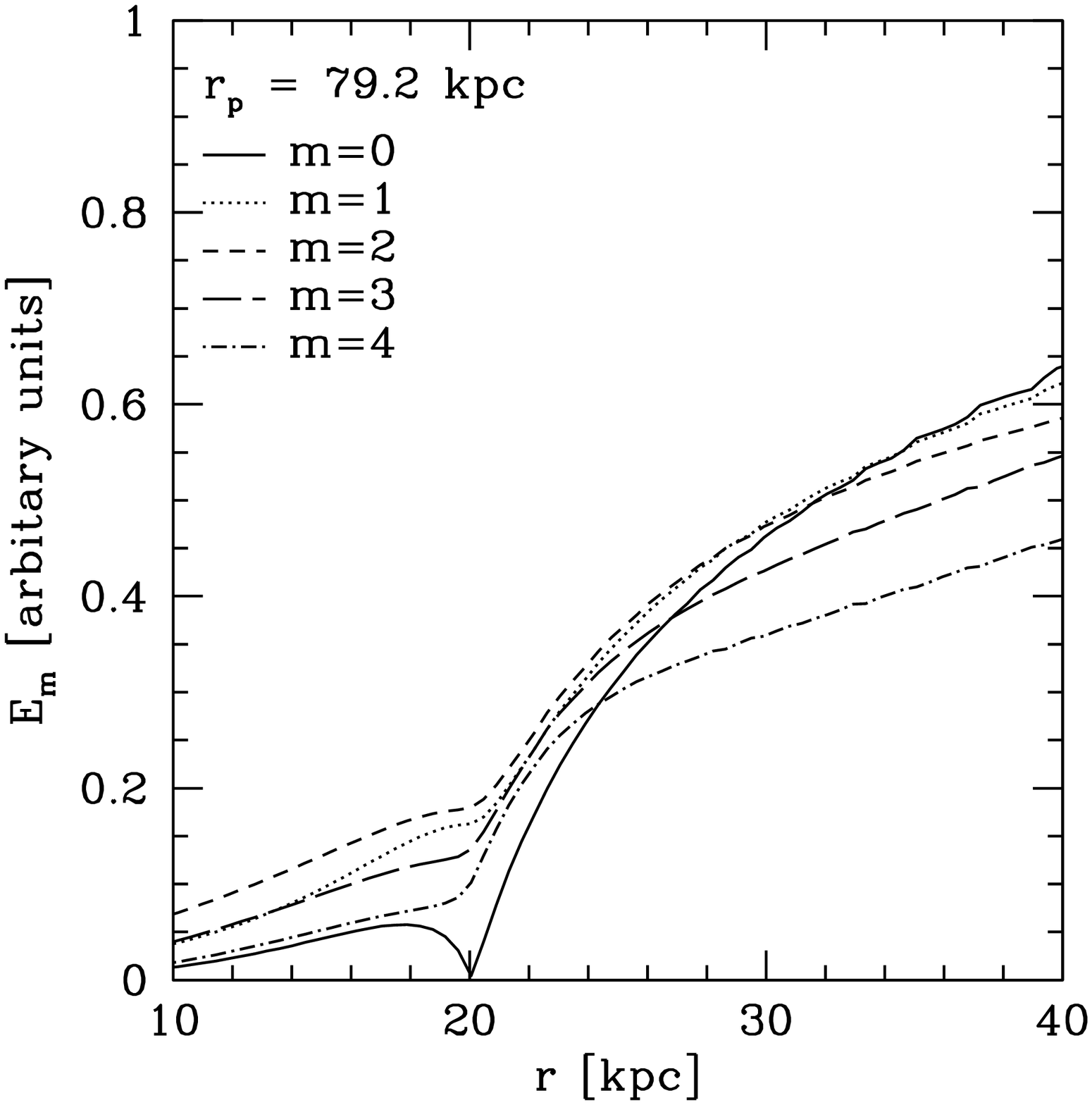}}
\end{center}
\caption{Modal energies of the disc in response to the disturbances by a coplanar 1:100 perturber with an $r_{\rm peri} = 20$ kpc.  In the upper left hand corner of each plot is the radial distance of the perturber.  Initially when the perturber is far away (a), only the other disc responds to the perturber.  However, as the perturber gets closer, more and more of the disc is perturbed.  In (c), the perturber is at $\rperi$ and we note a kink in the $m=0$ modal energy at $r=\rperi$.  This kink on the modal energy shows up in the subsequent plots (d), (e), and (f) in the other modes as well.  This kink is the result on an increase modal energy response at $\rperi$ and serves as an indicator of $\rperi$ of perturbing subhaloes.  We also note the advantage of studying the modal energy as modal energy is fixed after the interaction with the perturber in the absence of dissipation.}
\label{f:Em100_20kpc}
\end{figure*}

As we mentioned earlier in \S\ref{sec:equations}, the modal equations reduce to a number of simple harmonic oscillators.  A passing perturber excites these simple harmonic oscillators, each to a different amplitude and phase. In the absence of dissipation, the modal amplitude or energy remains constant.  We illustrate this point in Figure \ref{f:Em100_20kpc} where we show six snapshots of
the orbit of a coplanar subhalo with $r_{\rm peri} = 20$ kpc.  Initially, the energy is  zero, but rises dramatically after the disc suffers an encounter, before asymptoting to a constant.  This property of the modal energy response makes them illuminating for studying the response of the disc.  We should point out that while the energy is already substantial when the perturber is at pericenter, the {\it density} response of the disc does not happen until much later.  This is because, the energy imparted by the perturber at pericenter serves mainly to increase the perturbed velocity.  However, this perturbed velocity only translates into perturbed density after $\sim$ dynamical time.

\begin{figure*}
\begin{center}
\subfigure[]{
\includegraphics[width=.45\textwidth]{Em100_20kpc/5.eps}}
\subfigure[]{
\includegraphics[width=.45\textwidth]{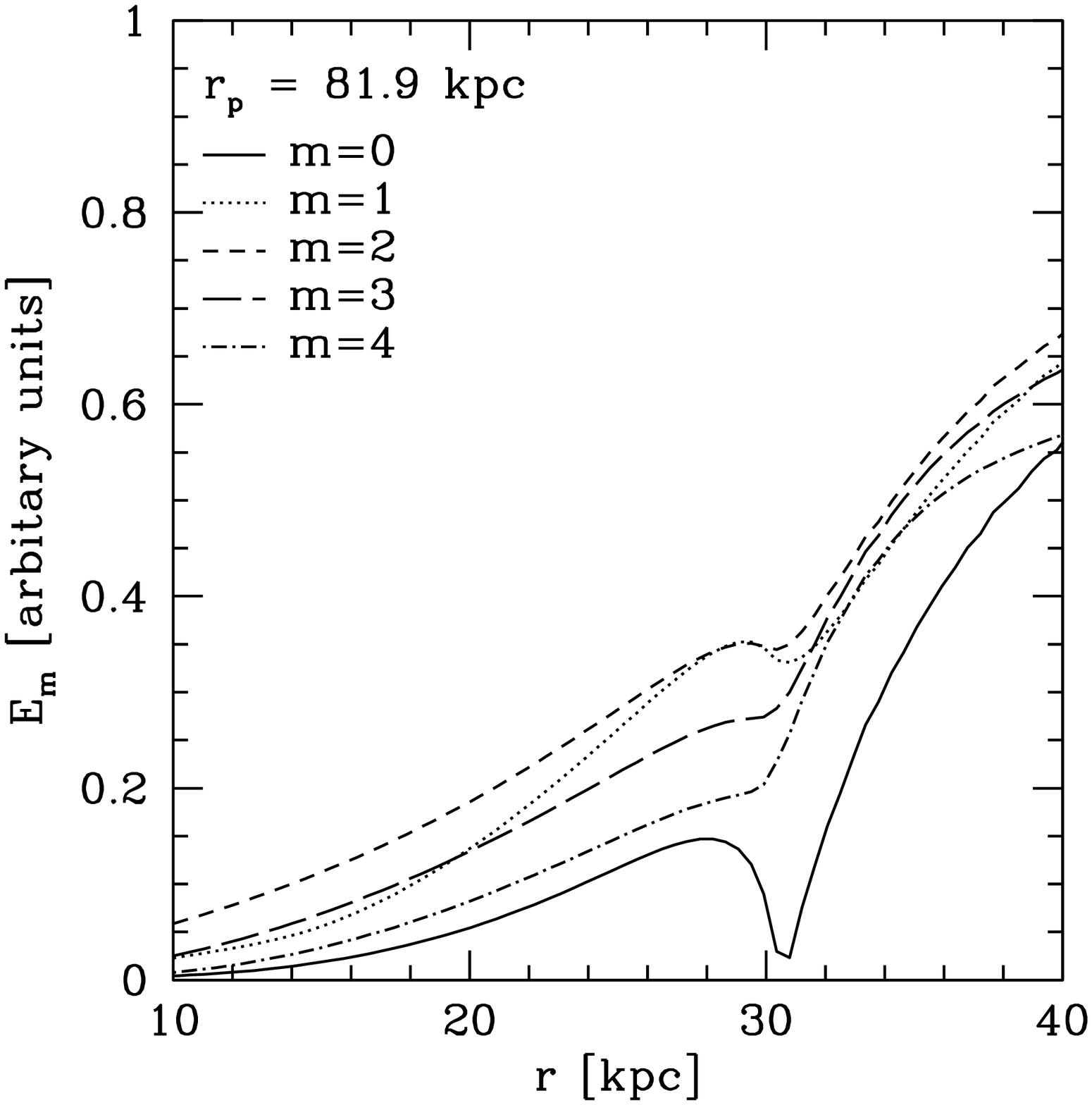}}
\end{center}
\caption{Modal energies of the disc for a 1:100 perturber with $\rperi = 20$ kpc (a) and $30$ kpc (b).  In both case the perturber sits at roughly $r_p\approx 80$ kpc after its interaction.  As the modal energy is fixed at this point in the absence of dissipation, we can look in detail at the structure of the interaction.  Note that in both cases, a kink on the modal energy appears at $r=\rperi$.  This strengthens the point made in Figure \ref{f:Em100_20kpc} that these kinks serves as a good proxy for $\rperi$.}
\label{f:Em100}
\end{figure*}

\begin{figure*}
\begin{center}
\subfigure[]{
\includegraphics[width=.45\textwidth]{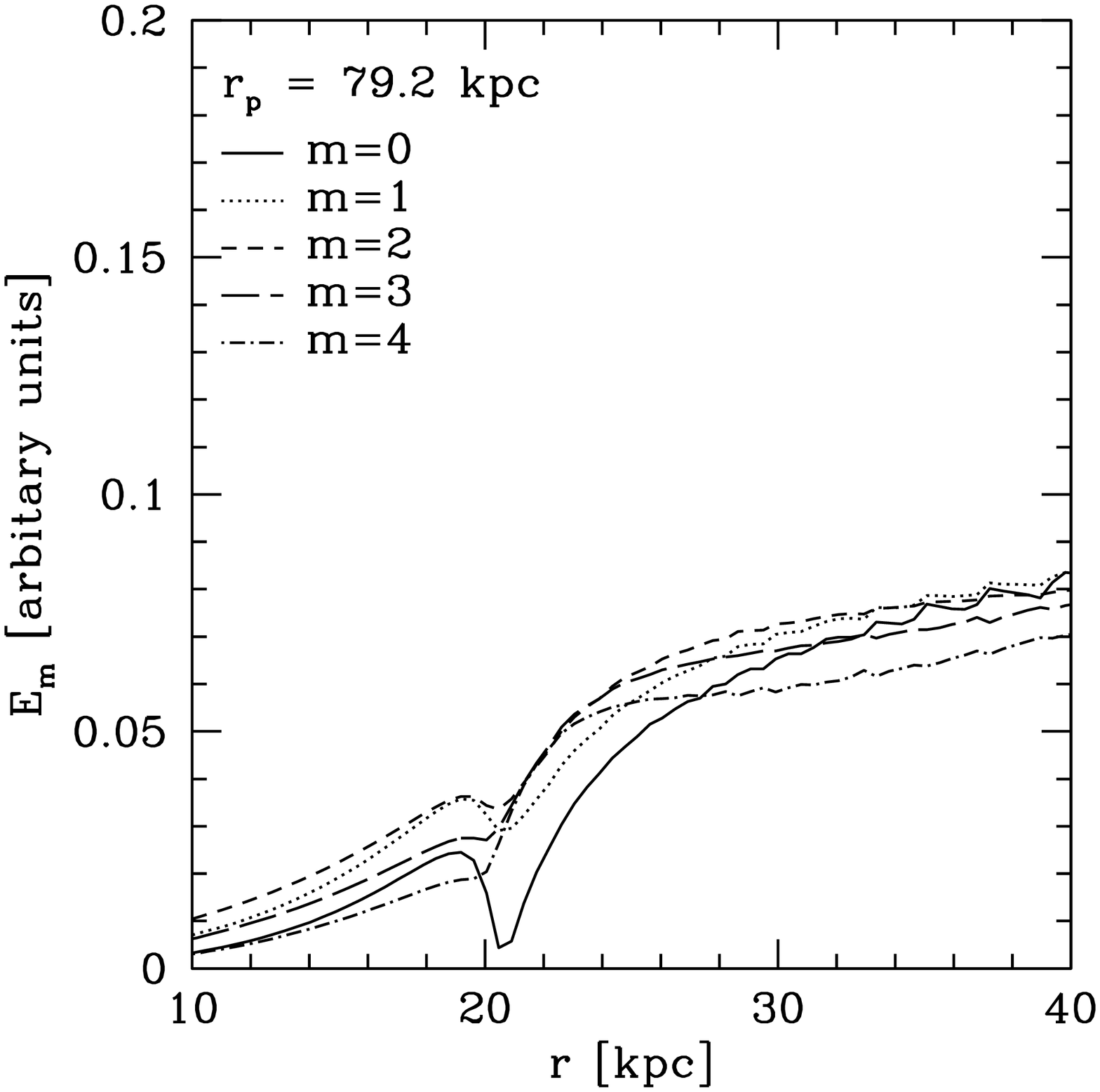}}
\subfigure[]{
\includegraphics[width=.45\textwidth]{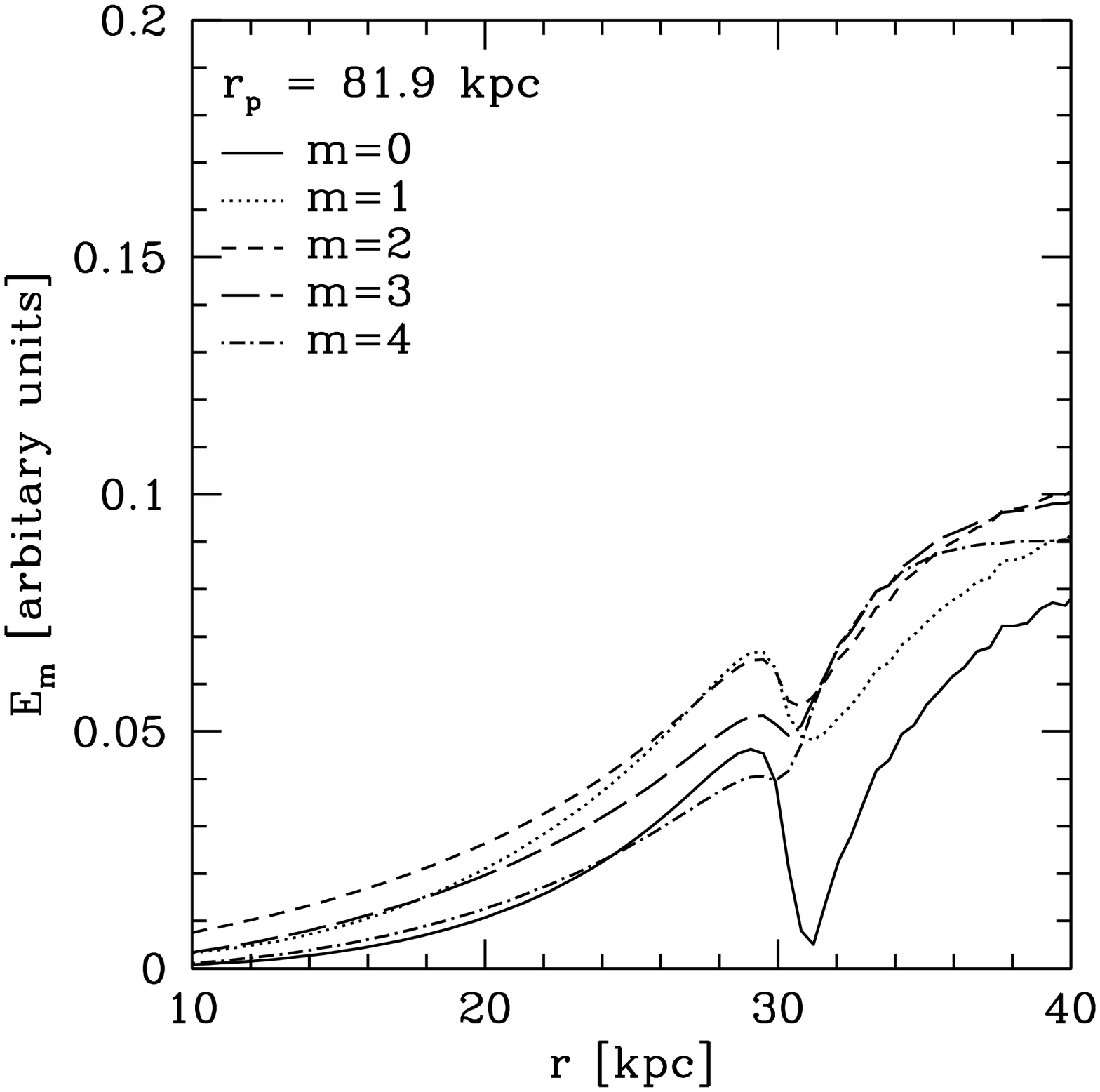}}
\end{center}
\caption{Same as Figure \ref{f:Em100} but for a 1:1000 perturber with $\rperi = 20$ kpc (a) and $30$ kpc (b).}
\label{f:Em1000}
\end{figure*}

How does the modal (energy) profile depend on the orbital parameter?  To answer this question, let us first fix the mass and the orbit to be that of a parabolic ($E=0$) coplanar orbit so that the modal profile only depends on the pericenter distance of the perturber. In Figure \ref{f:Em100}, we plot modal energies as a function of radius for a 1:100 encounter for $r_{\rm peri} = 10$ kpc (left plot) and $20$ kpc (right plot).  Note the dramatic increase in modal energy for the oscillator at $r=r_{\rm peri}$.  Indeed this increase in modal energy is seen for 1:10 and 1:1000 coplanar encounters as shown in Figure \ref{f:Em1000}. 
We also note that the amplitude of the modal profile depends only on the mass and is only weakly dependent on $\rperi$ at least at large distances.  

\begin{figure*}
\begin{center}
\subfigure[]{
\includegraphics[width=.23\textwidth]{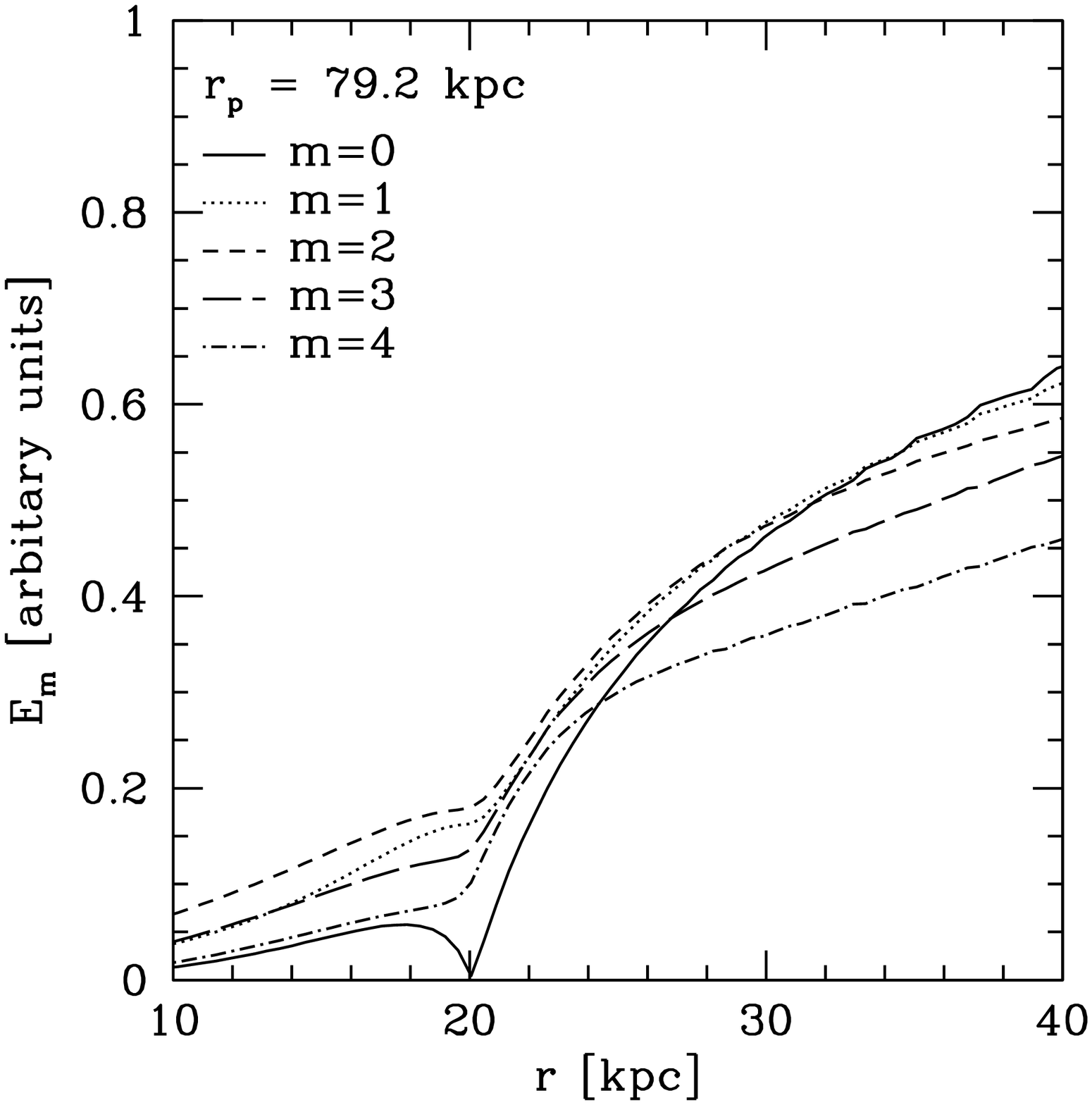}}
\subfigure[]{
\includegraphics[width=.23\textwidth]{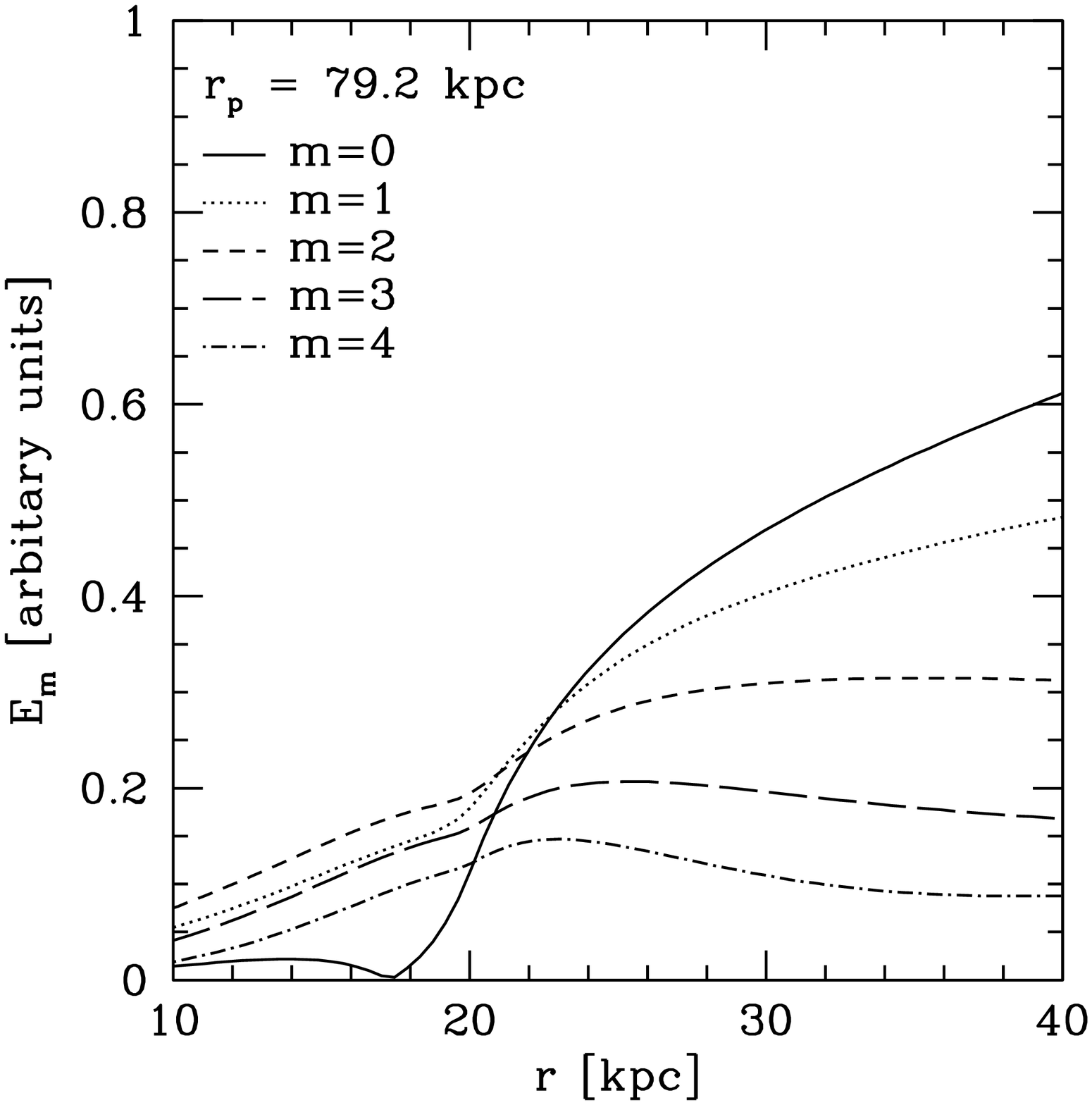}}
\subfigure[]{
\includegraphics[width=.23\textwidth]{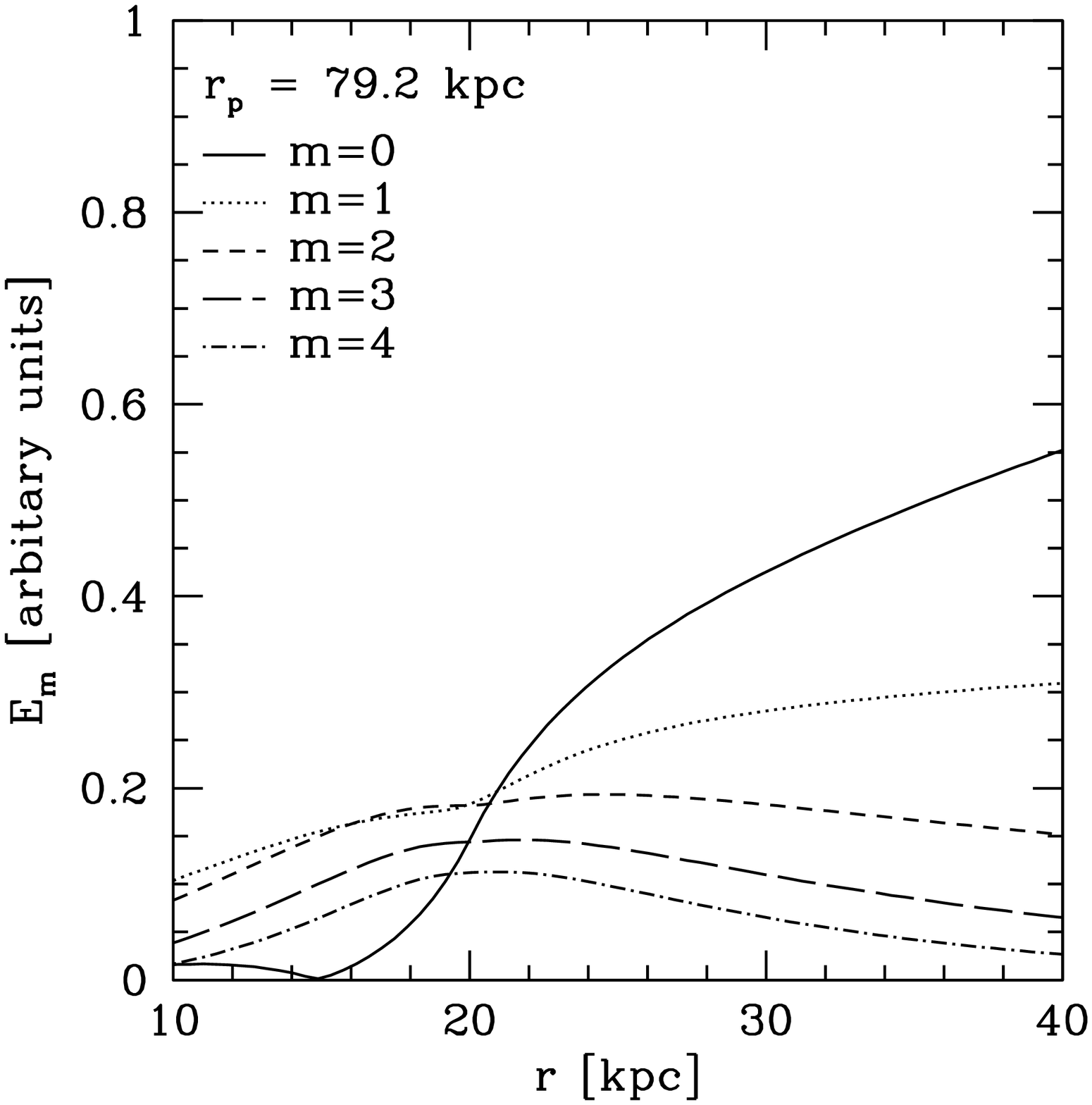}}
\subfigure[]{
\includegraphics[width=.23\textwidth]{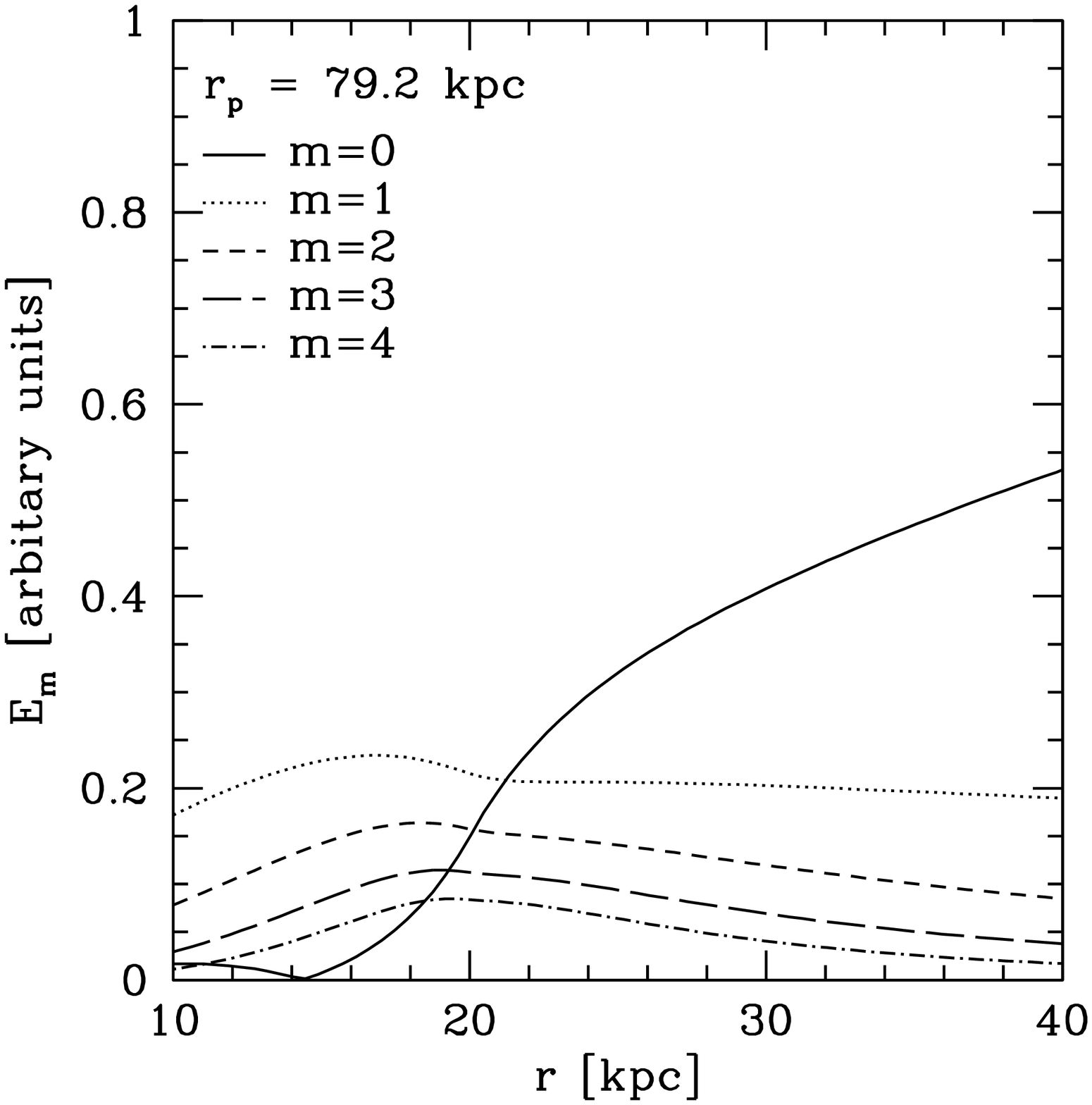}}
\subfigure[]{
\includegraphics[width=.23\textwidth]{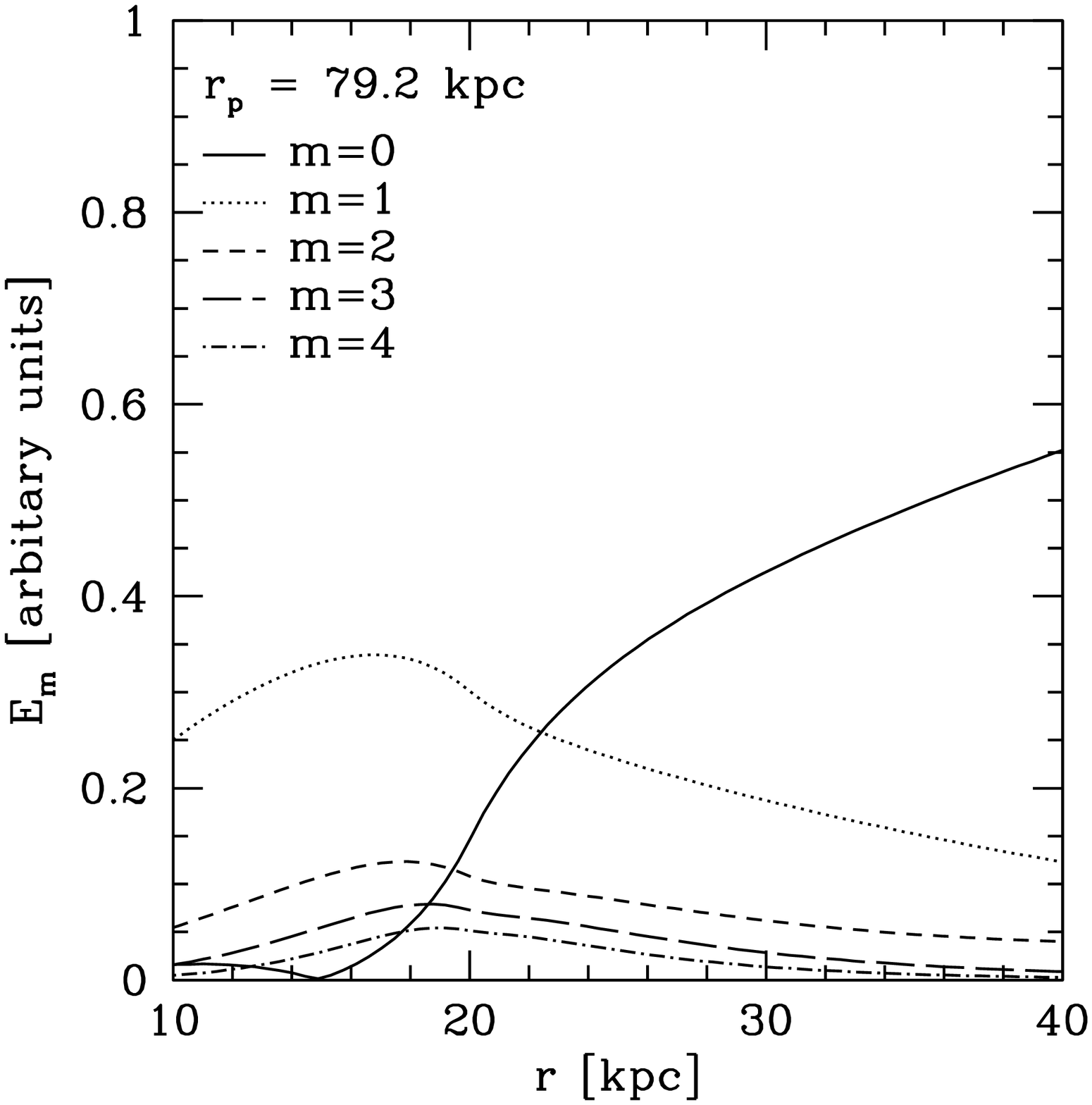}}
\subfigure[]{
\includegraphics[width=.23\textwidth]{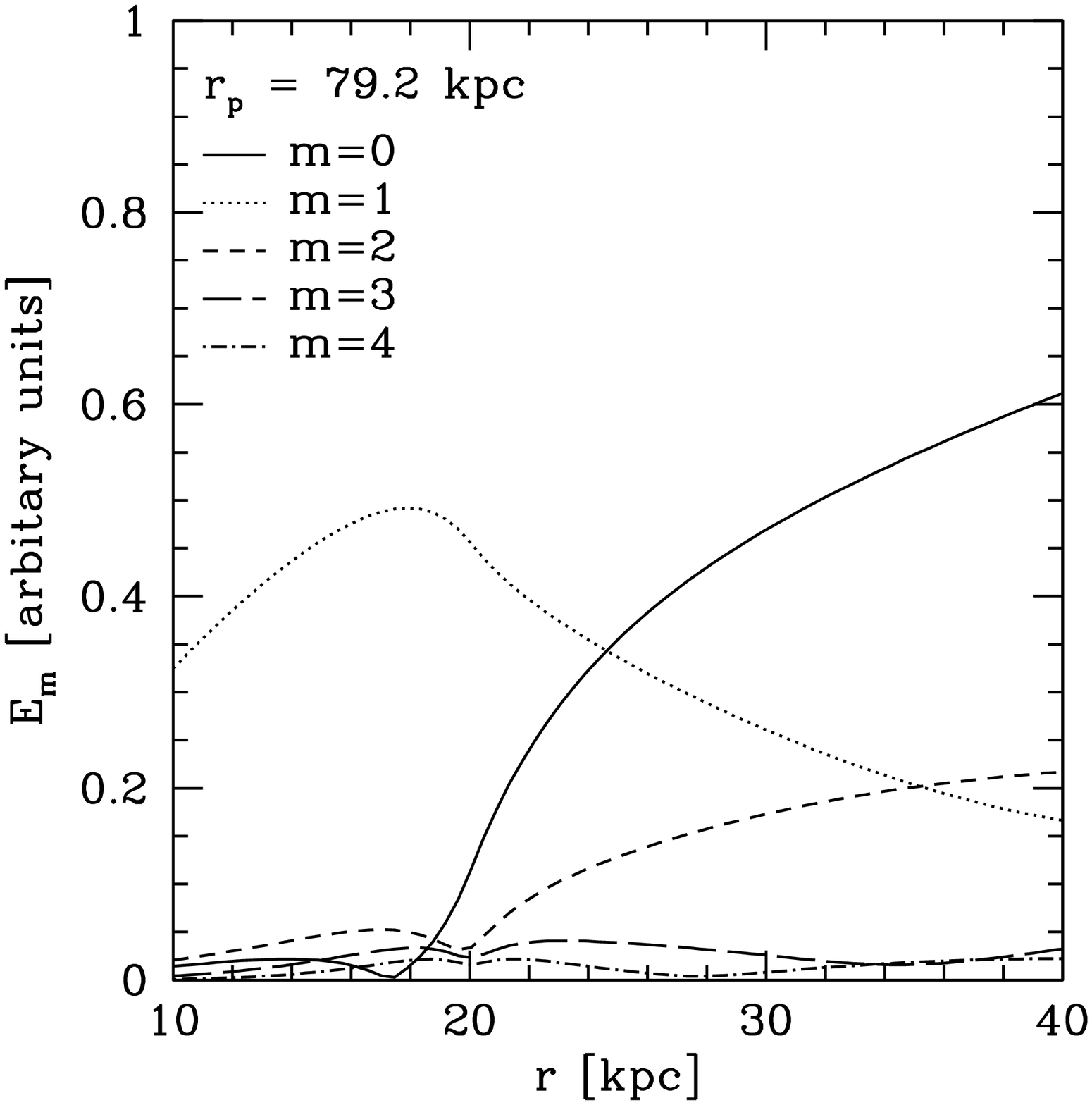}}
\subfigure[]{
\includegraphics[width=.23\textwidth]{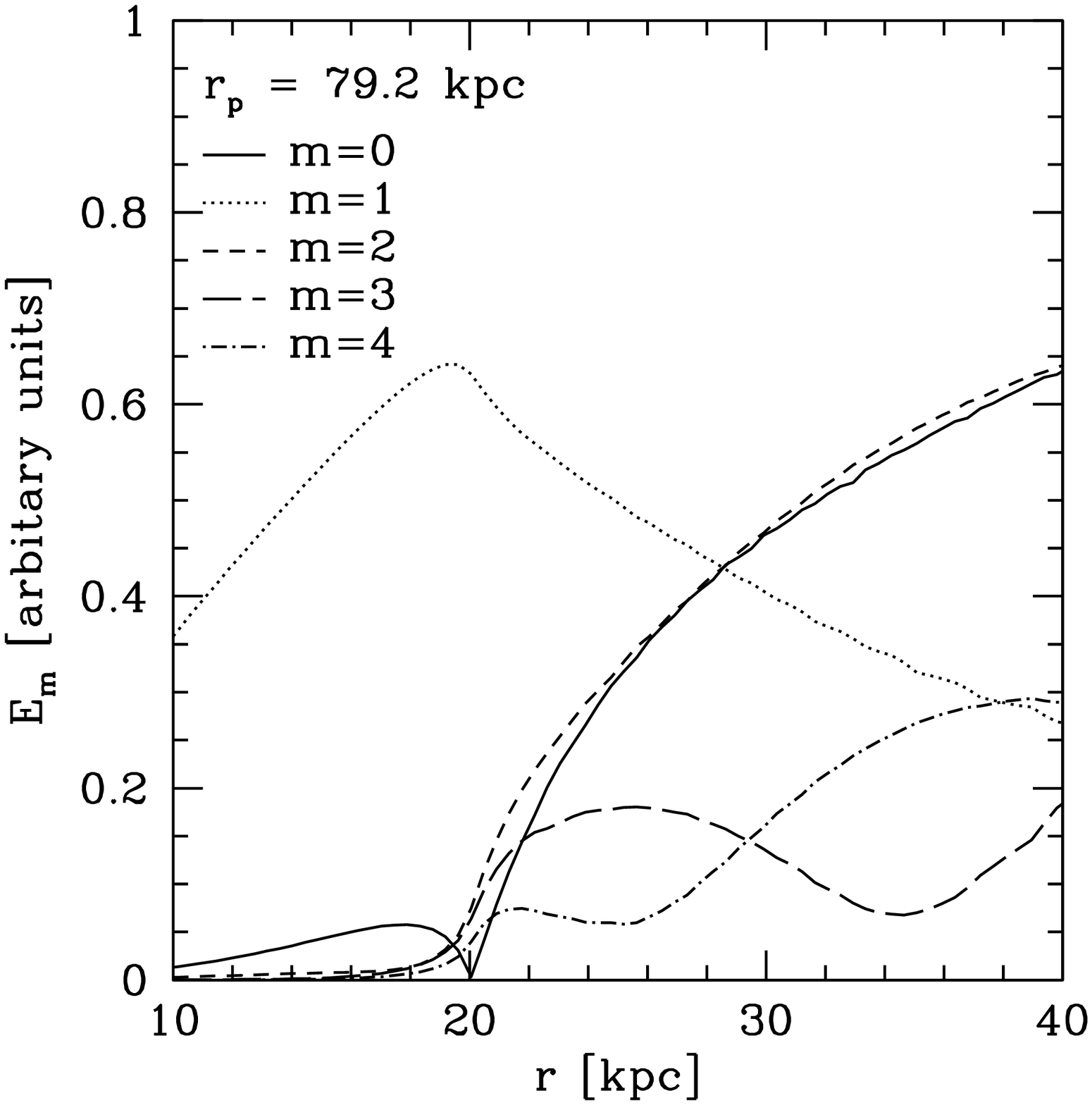}}
\end{center}
\caption{Modal energy profiles of a 1:100 perturber with $r_{\rm peri} = 20$ kpc for different
inclinations: $i = 90$ (a), 60 (b), 30 (c), 0 (d), -30 (e), -60 (f), -90 (g) degrees.  The kink in the modal energy profile in the coplanar case (a) and Figure \ref{f:Em100_20kpc} is only persistent for the $m=0$ mode for all inclinations. However, the modal energy profile near $r = \rperi$ shows features in all cases like peaks, e.g., $m=1$ case for -30 (e), -60 (f), and -90 (g) degrees. }
\label{f:Em100_inclination}
\end{figure*}

We now fix $r_{\rm peri} = 20$ kpc to study the effects of inclination in Figure \ref{f:Em100_inclination}. In the coplanar case studied above, all the modes showed a similar feature at $\rperi$.  Here, we instead see that only the $m=0$ mode consistently shows this same feature as the  coplanar case. Specifically, for all inclinations, ranging from $+\pi/2$ (prograde coplanar) to $0$ vertical to $-\pi/2$ (retrograde coplanar), the $m=0$ modal energy increases dramatically as $r > r_{\rm peri}$.    

The effect of inclination on the $m\neq 0$ modal energies is more complicated. 
For the coplanar prograde interaction, we also see a large increase in the modal
energy at $r=\rperi$ for all $m\neq 0$ modes.  However, this is now true for the
other interactions.  For instance, the retrograde coplanar interaction
($-\pi/2$) shows a large amplitude for the $m=1$ mode inside of $\rperi$ and it
decreases outward.  Nevertheless, it is clear the modal energies does show a
sudden change at $\rperi$. 

In these examples above, we have studied the effect of varying the inclination for purely prograde encounter, i.e., the case where the disc crossing radius is equal to the pericenter distance.  Though we do not present them here, we have also studied the case of fixing $\rperi$ and the inclination and varying the angle at which the perturber crosses the disc, i.e., $\phi \ne 0$ for the disc crossing point on the y-axis.  In these examples, the modal energies also show a change at $\rperi$, which suggest that looking for changes in the modal energy in the disc is a good way to deduce $\rperi$.  We briefly comment on this in the next subsection, but reserve a more extensive discussion on the operational details to future work.

\subsection{Density Response}

The modal energies provide a good proxy for the response of the disc, but our analysis of modal energies here is new.
We now link this modal energy response to the density response, which is the observed manifestation of these perturbations. To calculate the density response from the modal energies, we reconstruct the position of the particles using equations (\ref{eq:reconstruction 1}) - (\ref{eq:reconstruction 2}).  From this, we obtain the density of particles as a function of $r$, $\phi$, and $t$, from which the Fourier amplitudes can be calculated.  

\begin{figure*}
\begin{center}
\subfigure[]{
\includegraphics[width=.27\textwidth]{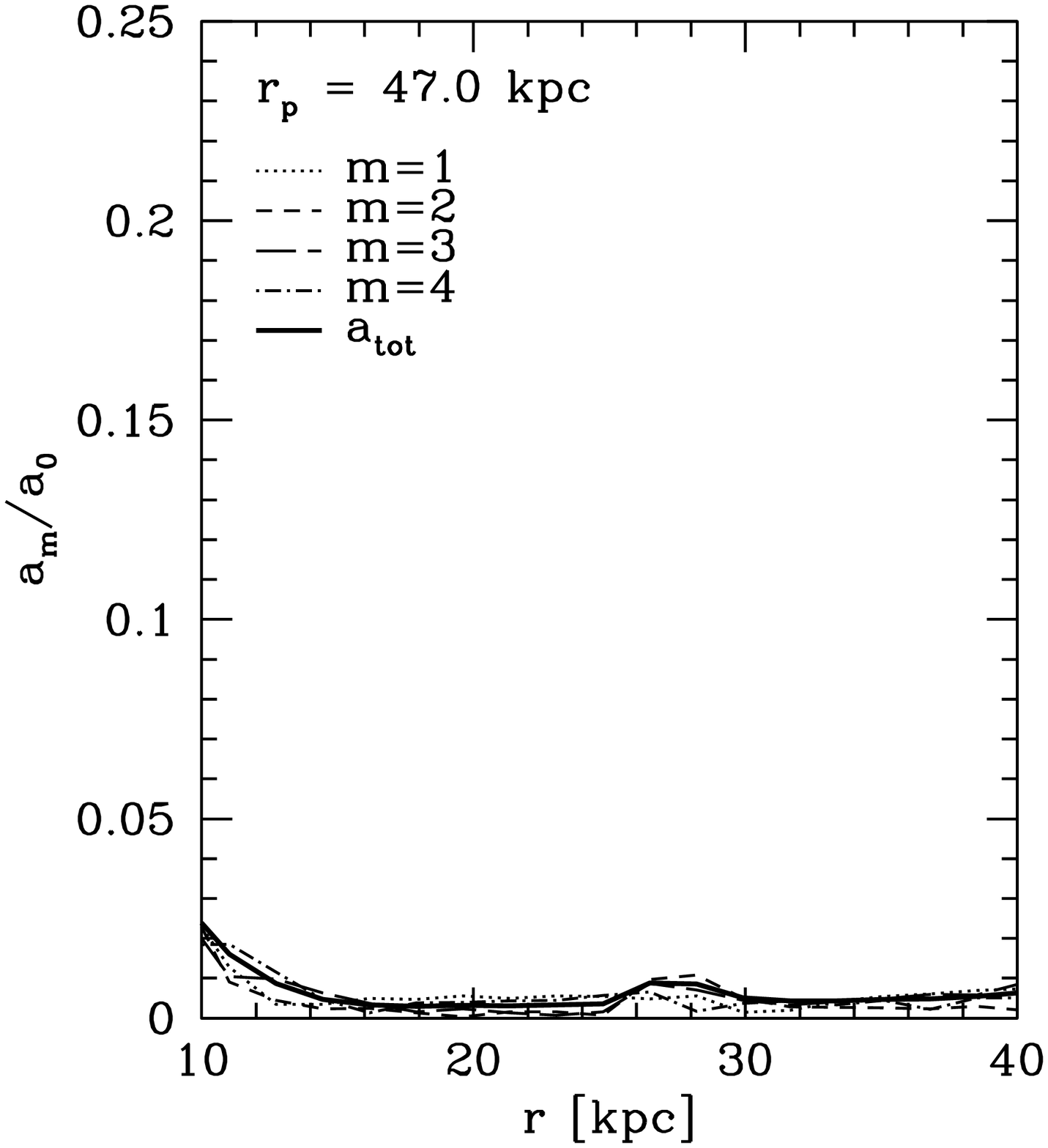}}
\subfigure[]{
\includegraphics[width=.27\textwidth]{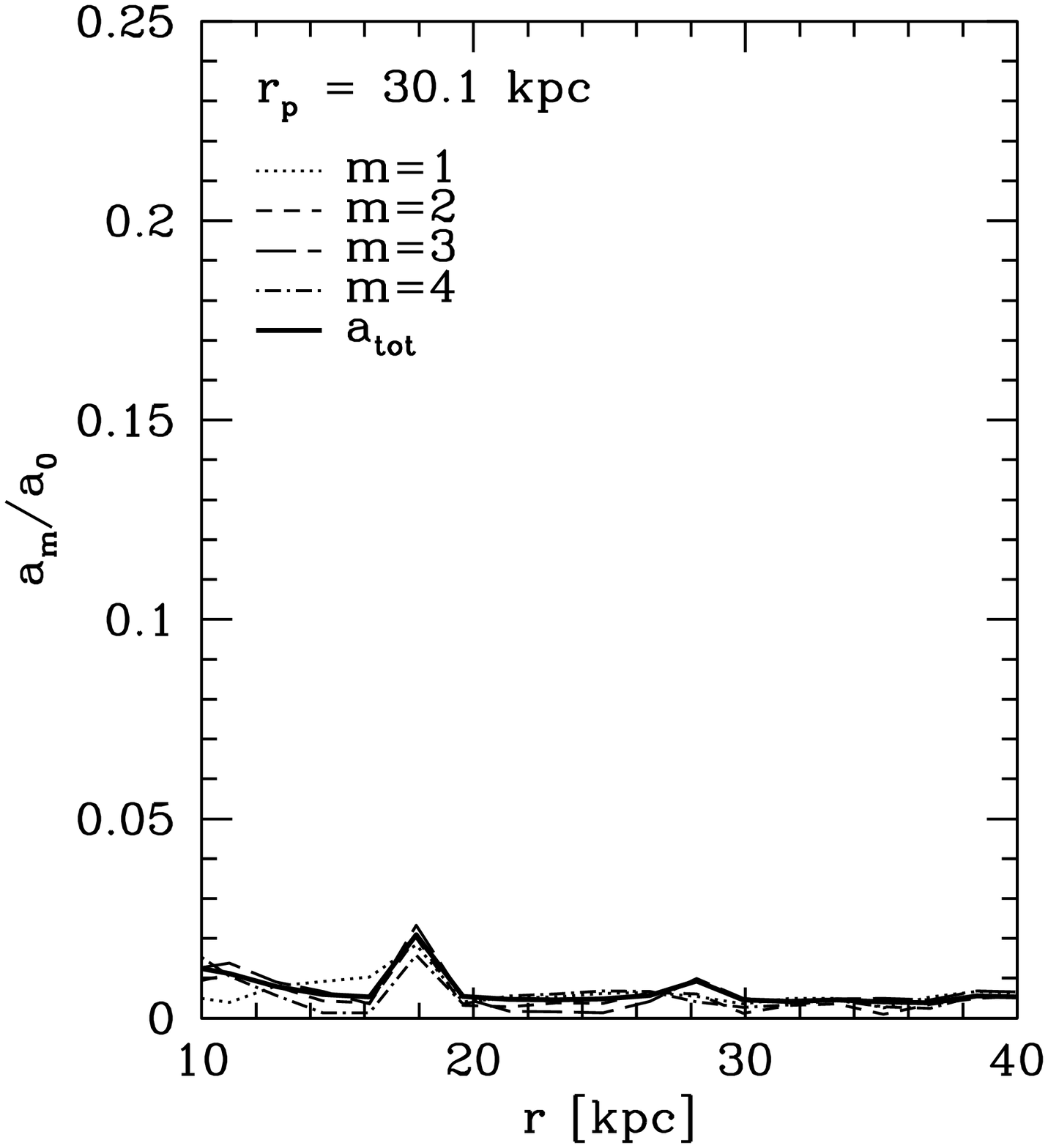}}
\subfigure[]{
\includegraphics[width=.27\textwidth]{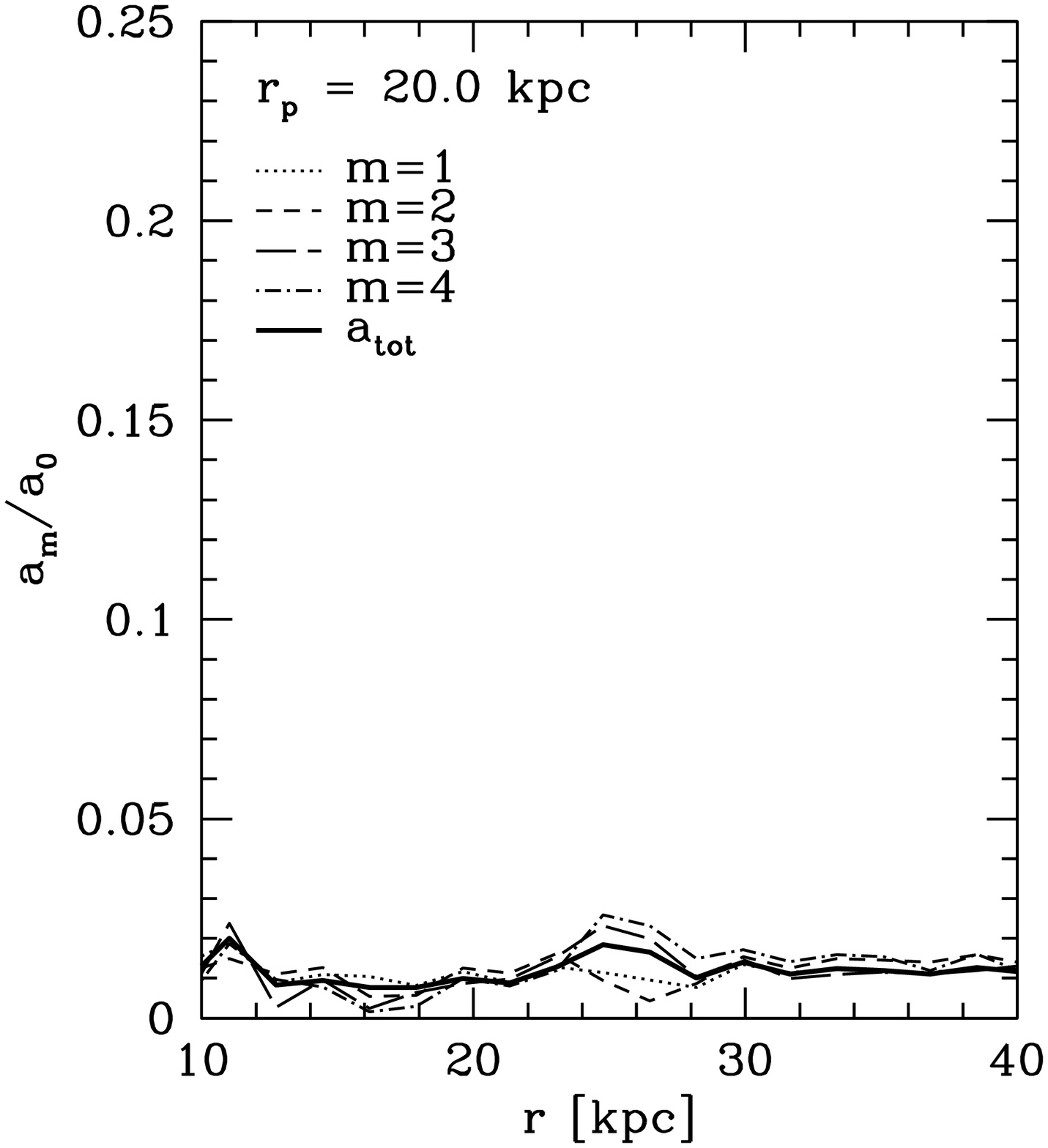}}
\subfigure[]{
\includegraphics[width=.27\textwidth]{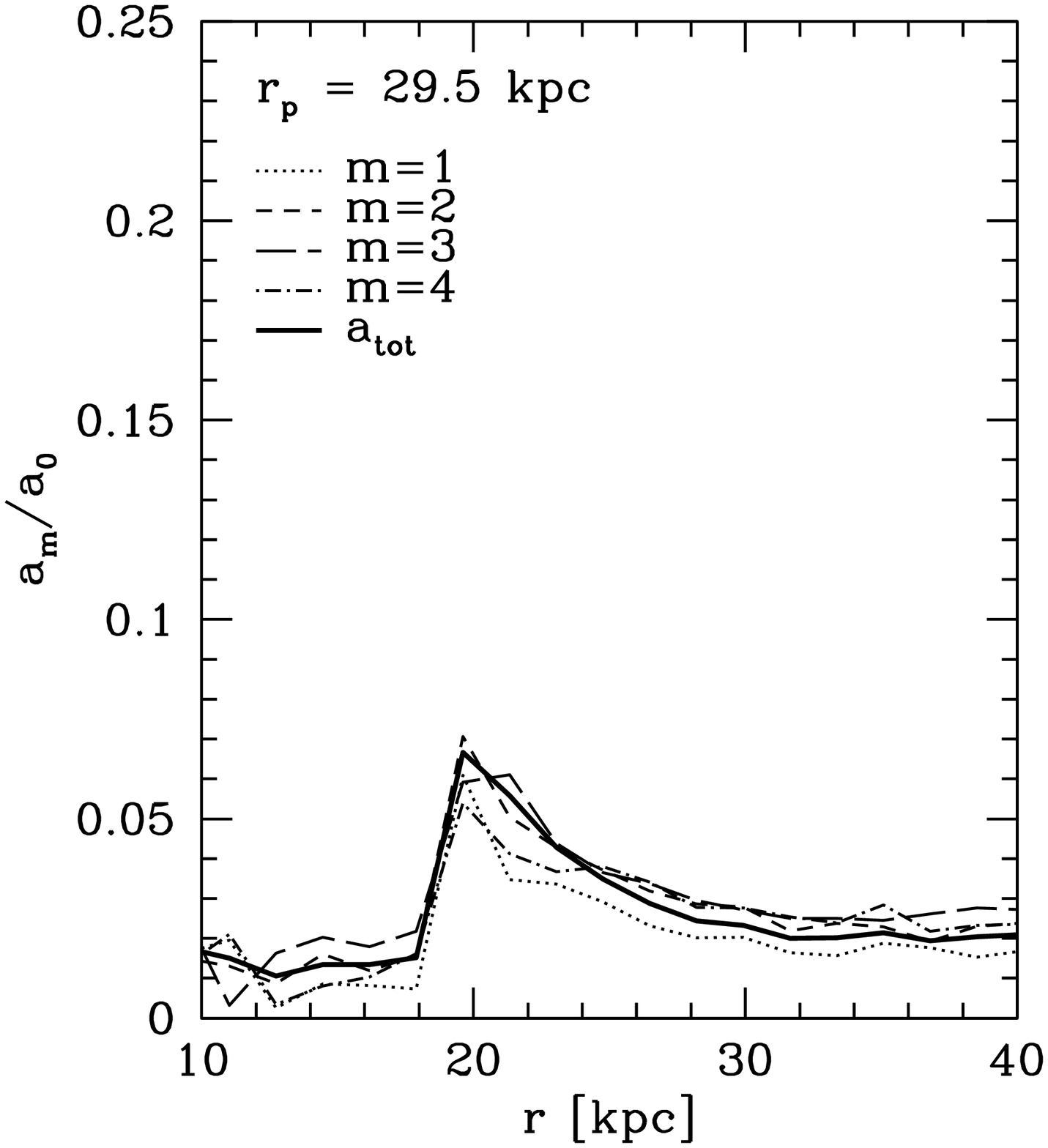}}
\subfigure[]{
\includegraphics[width=.27\textwidth]{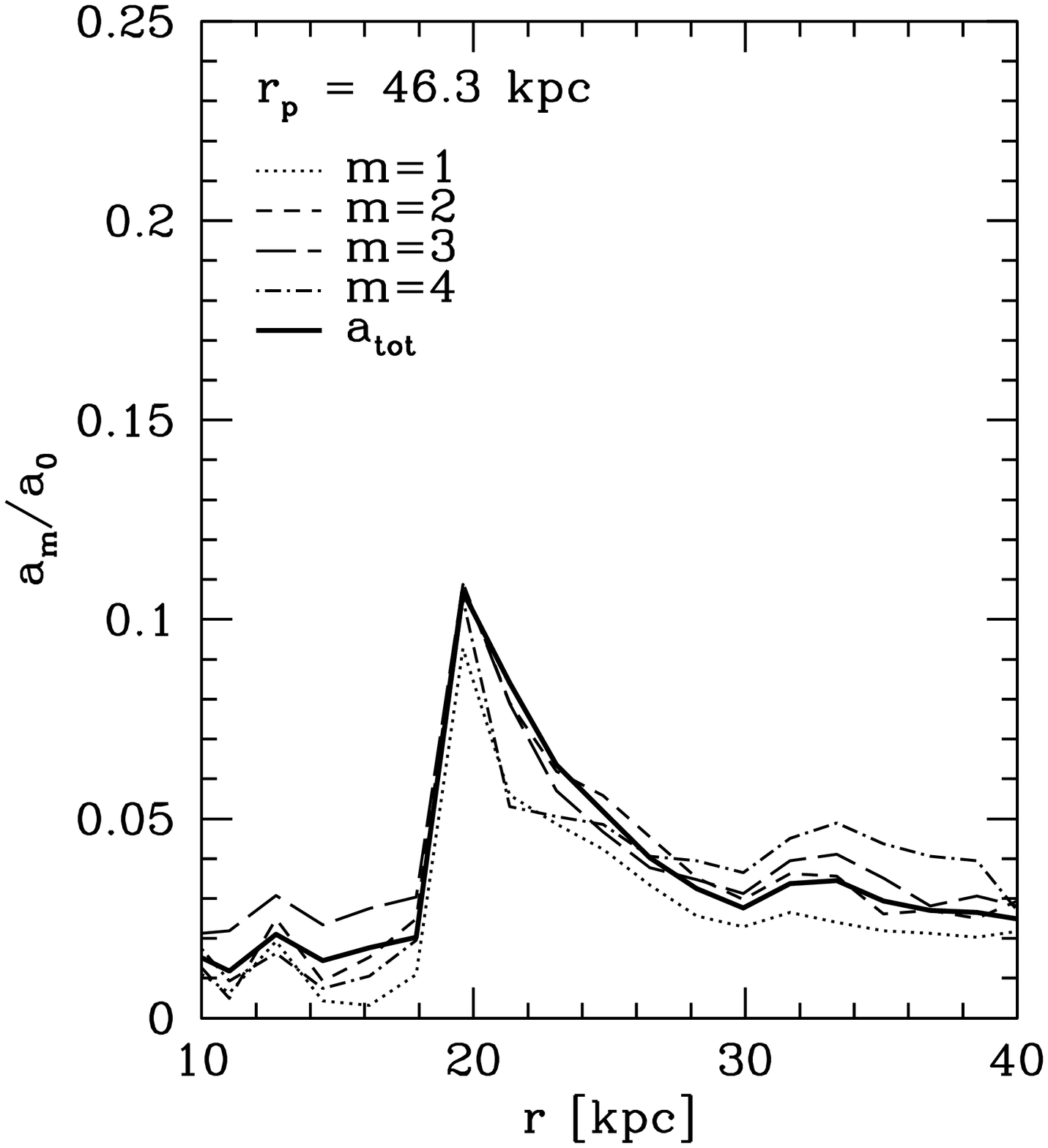}}
\subfigure[]{
\includegraphics[width=.27\textwidth]{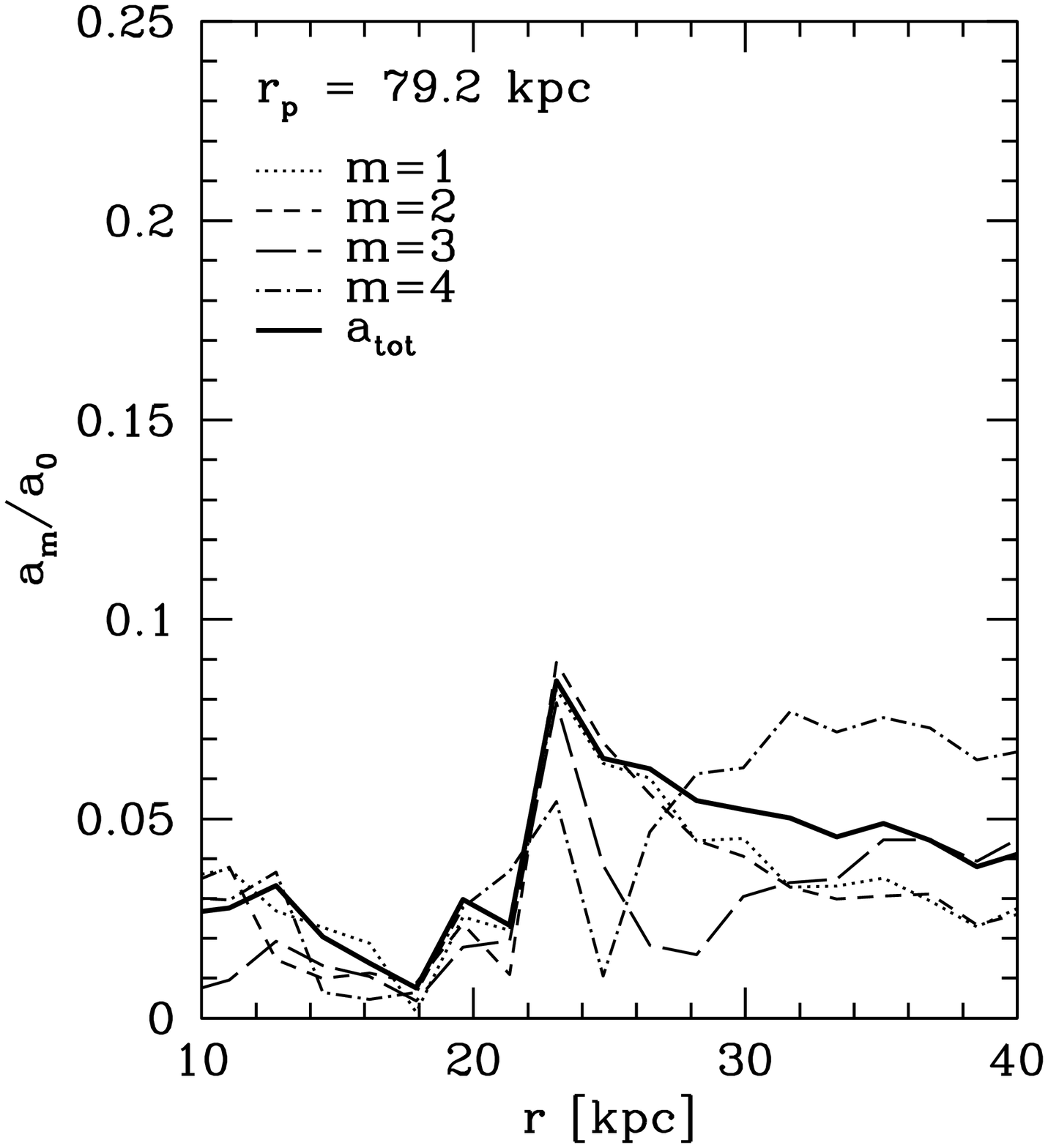}}
\end{center}
\caption{Fourier amplitudes of the density response of a disc suffering a 1:100 perturber encounter with $r_{\rm peri} = 20$ kpc. The perturber has a coplanar orbit ($i = 90$).  This corresponds to the modal energy calculation of Figure \ref{f:Em100_20kpc}, where we showed that the encounter generates a kink in the modal energy at $r=\rperi$.  Here we see that the result of this modal energy kink is reflected in the density response of the disc in (d), (e), and (f).  Namely, the density response is peaked at $r\approx \rperi$.  The response of $m=1-4$ is compared to the
$m=0$ mode.  For the $m=0$ line shown, we have only included the perturbation and compared it to the initial uniform background. }
\label{f:dens100_20kpc}
\end{figure*}

Our analysis of modal energies show that they are localized to certain regions of the disc.  This suggests that the density perturbations also follow a similar pattern.  We show this in Figure \ref{f:dens100_20kpc} where we show density perturbation from modal theory for the same six snapshots of the orbit of a coplanar subhalo with $r_{\rm peri} = 20$ kpc as in Figure \ref{f:Em100_20kpc}.  Note that we have only not included the $m=0$ mode here in constrast to the modal energy comparison.  While, we can define such an axisymmetric perturbation based on the initial (unperturbed) configuration, such a construction would not be well defined if we compare this to observations or simulations (see \S\ref{sec:sph}).  As we anticipate such a comparison, we have ignored the perturbed $m=0$ mode in the density response.  In addition, we have normalized the $m=1-4$ modes relative to the {\it complete} $m=0$ mode, which is dominated by the unperturbed component to further facilitate comparisons to observations and simulations.  

\begin{figure*}
\begin{center}
\subfigure[]{
\includegraphics[width=.27\textwidth]{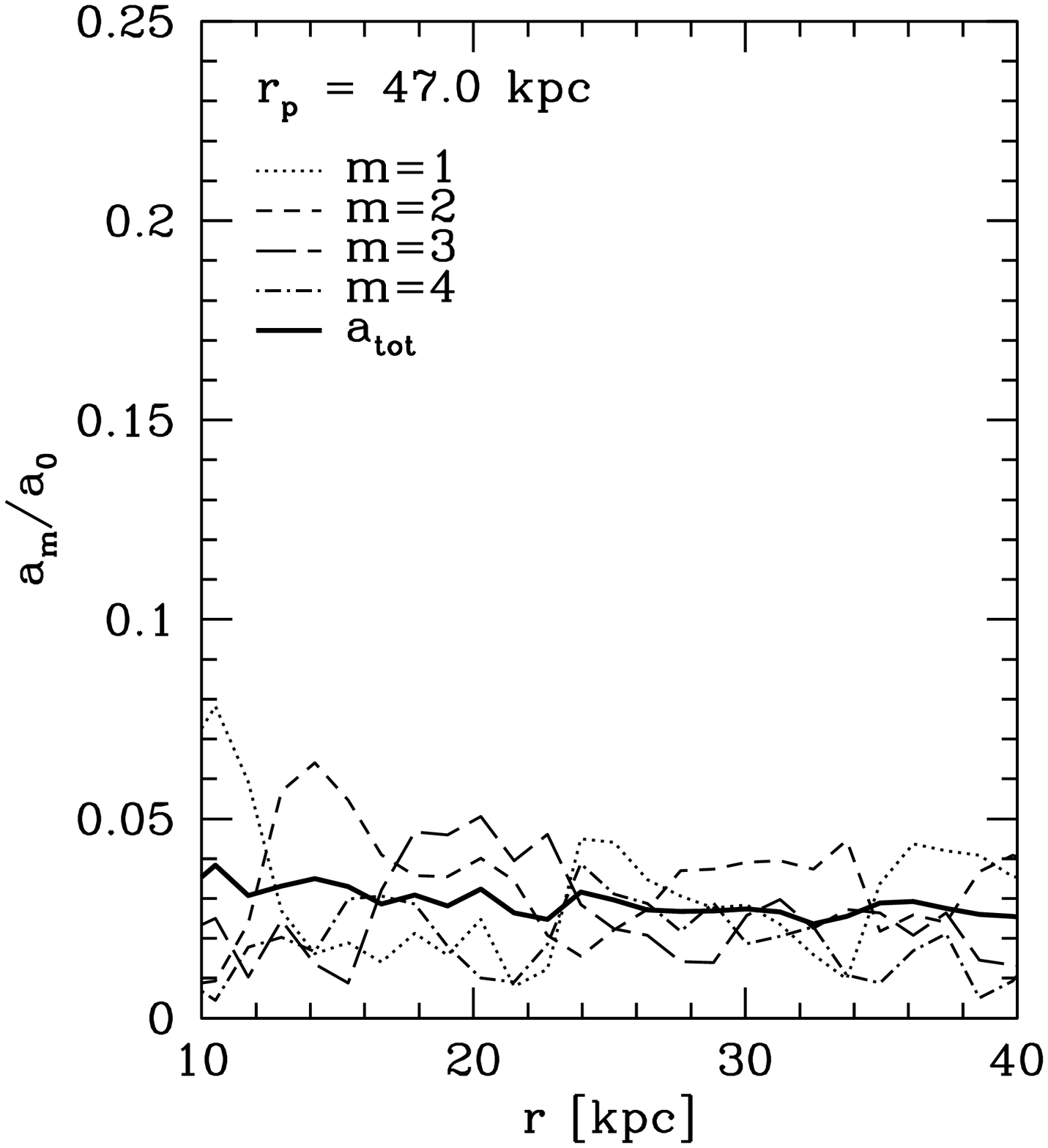}}
\subfigure[]{
\includegraphics[width=.27\textwidth]{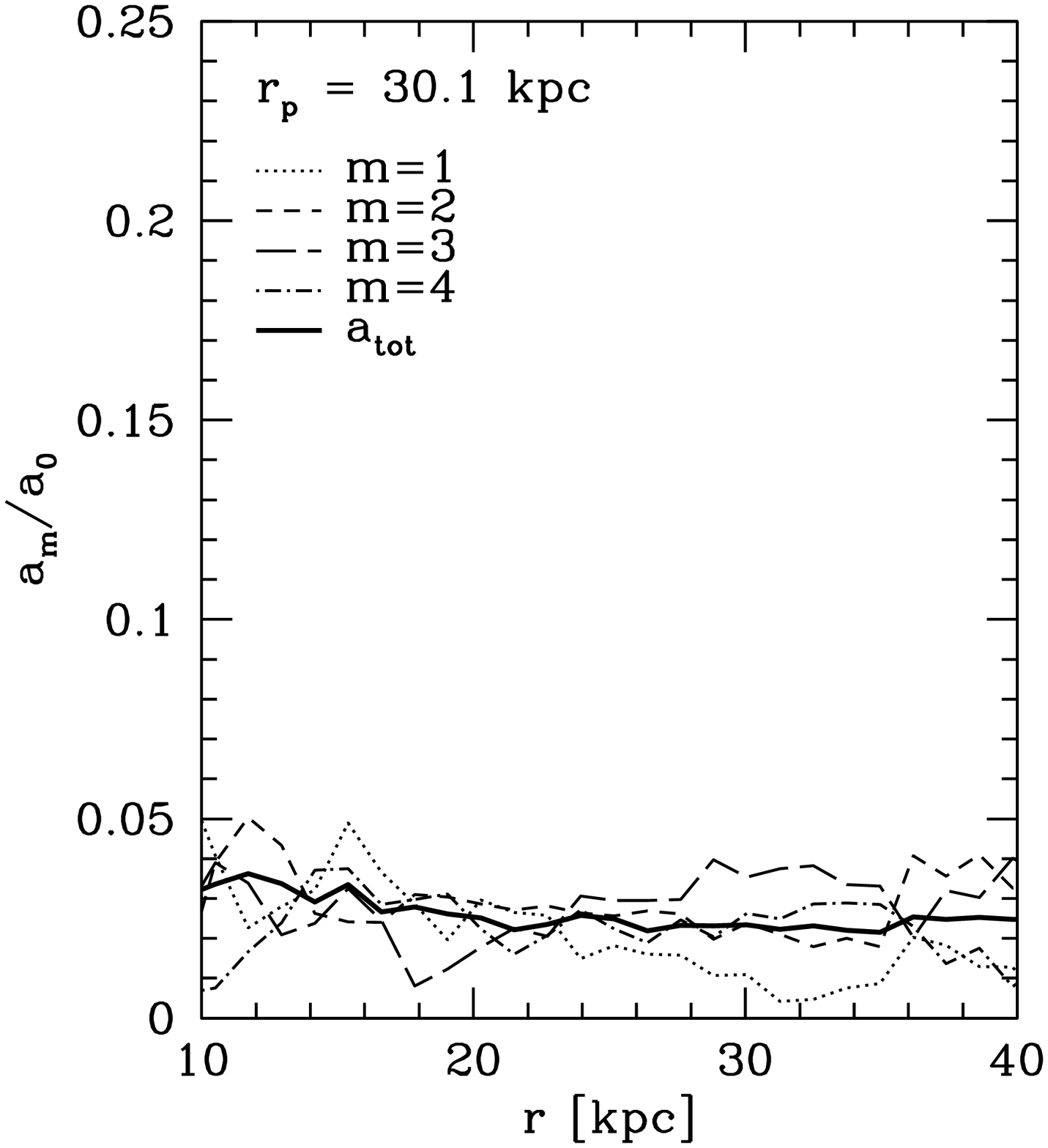}}
\subfigure[]{
\includegraphics[width=.27\textwidth]{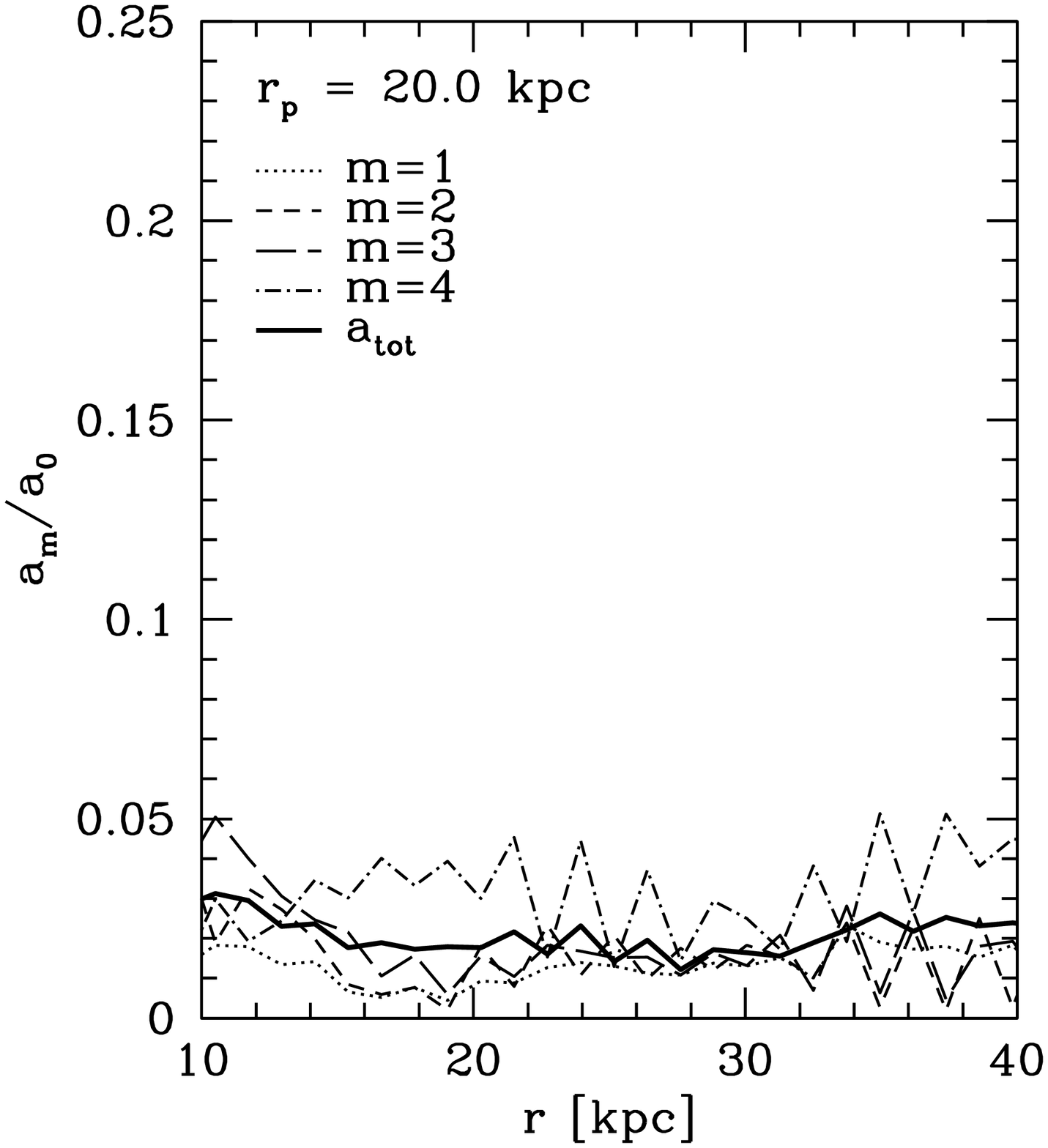}}
\subfigure[]{
\includegraphics[width=.27\textwidth]{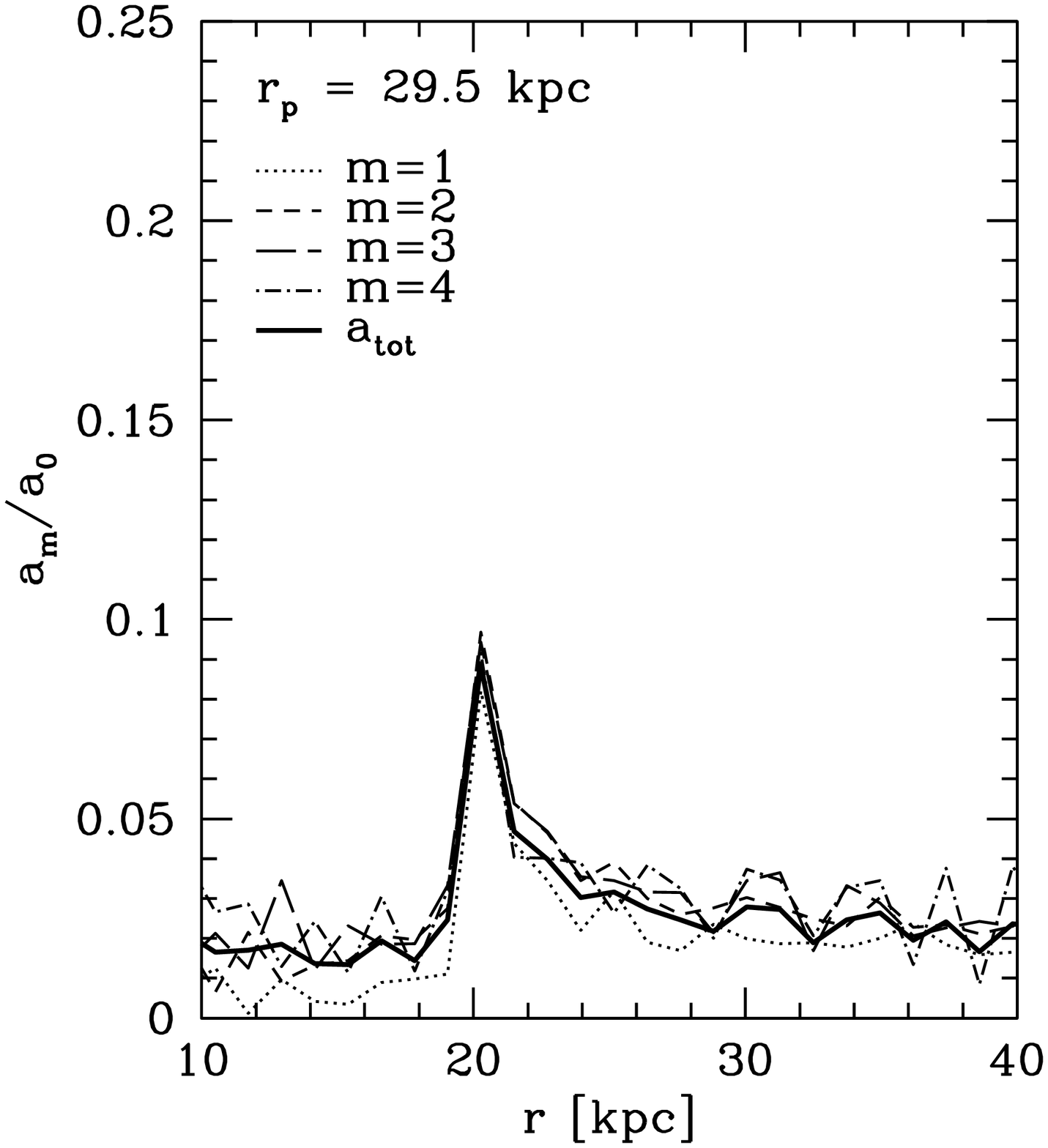}}
\subfigure[]{
\includegraphics[width=.27\textwidth]{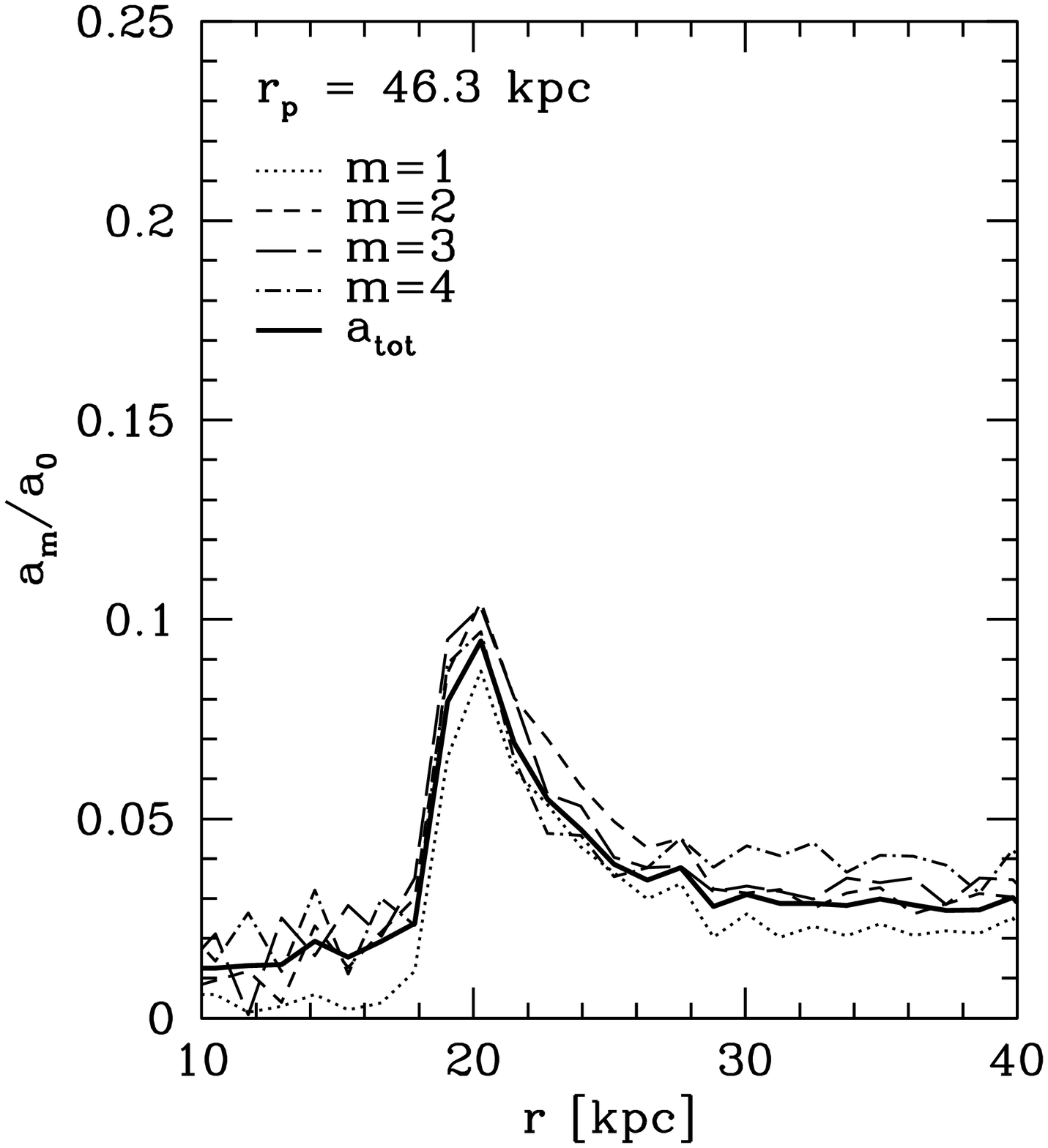}}
\subfigure[]{
\includegraphics[width=.27\textwidth]{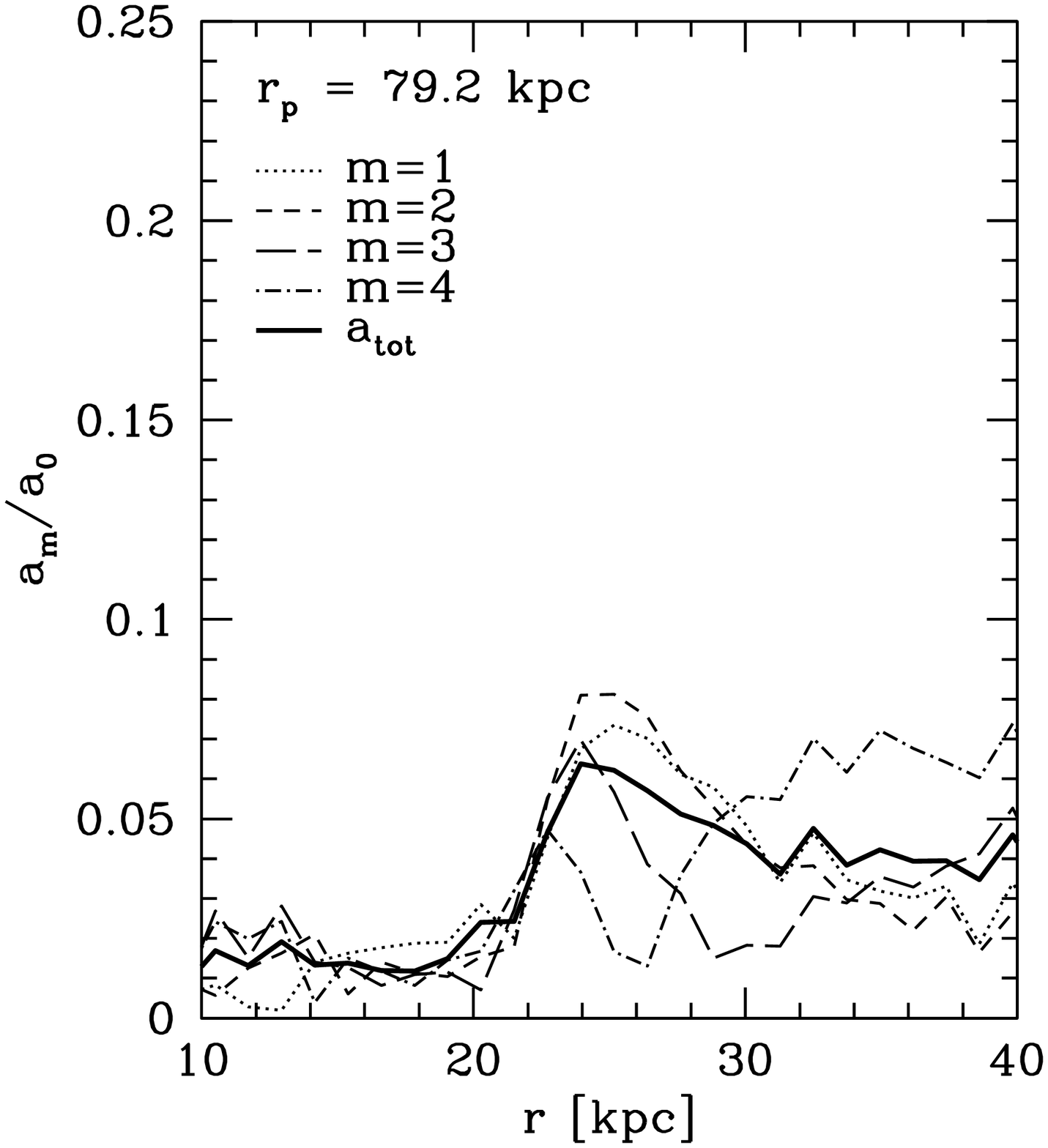}}
\end{center}
\caption{Same as Figure \ref{f:dens100_20kpc} but using the test particle
calculation.}
\label{f:dens100_20kpc_nbody}
\end{figure*}

We also show the same six snapshots using a full test-particle calculation in Figure \ref{f:dens100_20kpc_nbody}.  Note the similarity of Figure \ref{f:dens100_20kpc} and \ref{f:dens100_20kpc_nbody}, which increases our confidence in our method of solution.  Comparing Figure \ref{f:Em100_20kpc} and \ref{f:dens100_20kpc} side by side, it is clear that the modal energies are already significant as the perturber crossed $\rperi$, while the density perturbations occur later.  As discussed above, this is because the modal energies is initially concentrated in the perturbed velocities and the densities response only arises after $\sim$ dynamical time.

The abrupt increase in the modal energies at $r=\rperi$, which was previously identified by looking at modal energies, is also seen in the density response.  The density response like the modal energies remains small in the regions $r\lesssim \rperi$, but increase rapidly for $r\gtrsim \rperi$.  What is also interesting is that the identification of $\rperi$ from the density response in Figure \ref{f:dens100_20kpc} and \ref{f:dens100_20kpc_nbody} seems to be clearer than the modal energy response plotted in Figure \ref{f:Em100_20kpc}.  In addition, the density response appears to be much more localized than the modal energy response (Fig. \ref{f:Em100_20kpc}). Namely, it is strong between \rperi\ and $\rperi + 10$ kpc, whereas the modal energy response is strong all the way out to the end of the disc.

To help illiustrate the effect of this density response, we define a synthetic measure of the power of these modes as: 
\begin{equation}\label{eq:atot}
a_{\rm tot}(r) =  \sqrt{\sum_{m=0}^{4} \frac {|a_m(r)|^2} 5},
\end{equation}
and plot this in Figures \ref{f:dens100_20kpc} and \ref{f:dens100_20kpc_nbody}.

\begin{figure*}
\begin{center}
\subfigure[]{
\includegraphics[width=.23\textwidth]{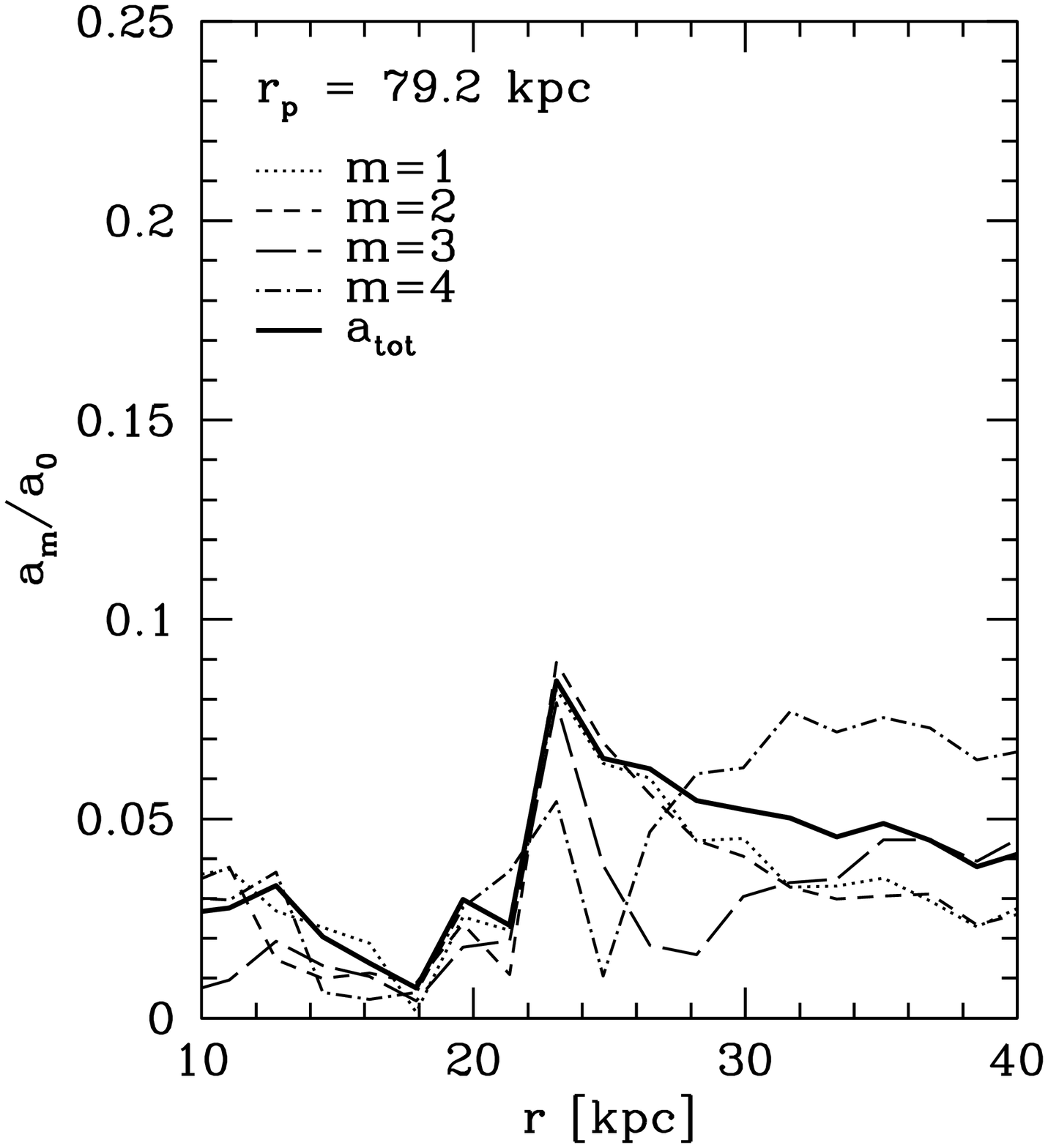}}
\subfigure[]{
\includegraphics[width=.23\textwidth]{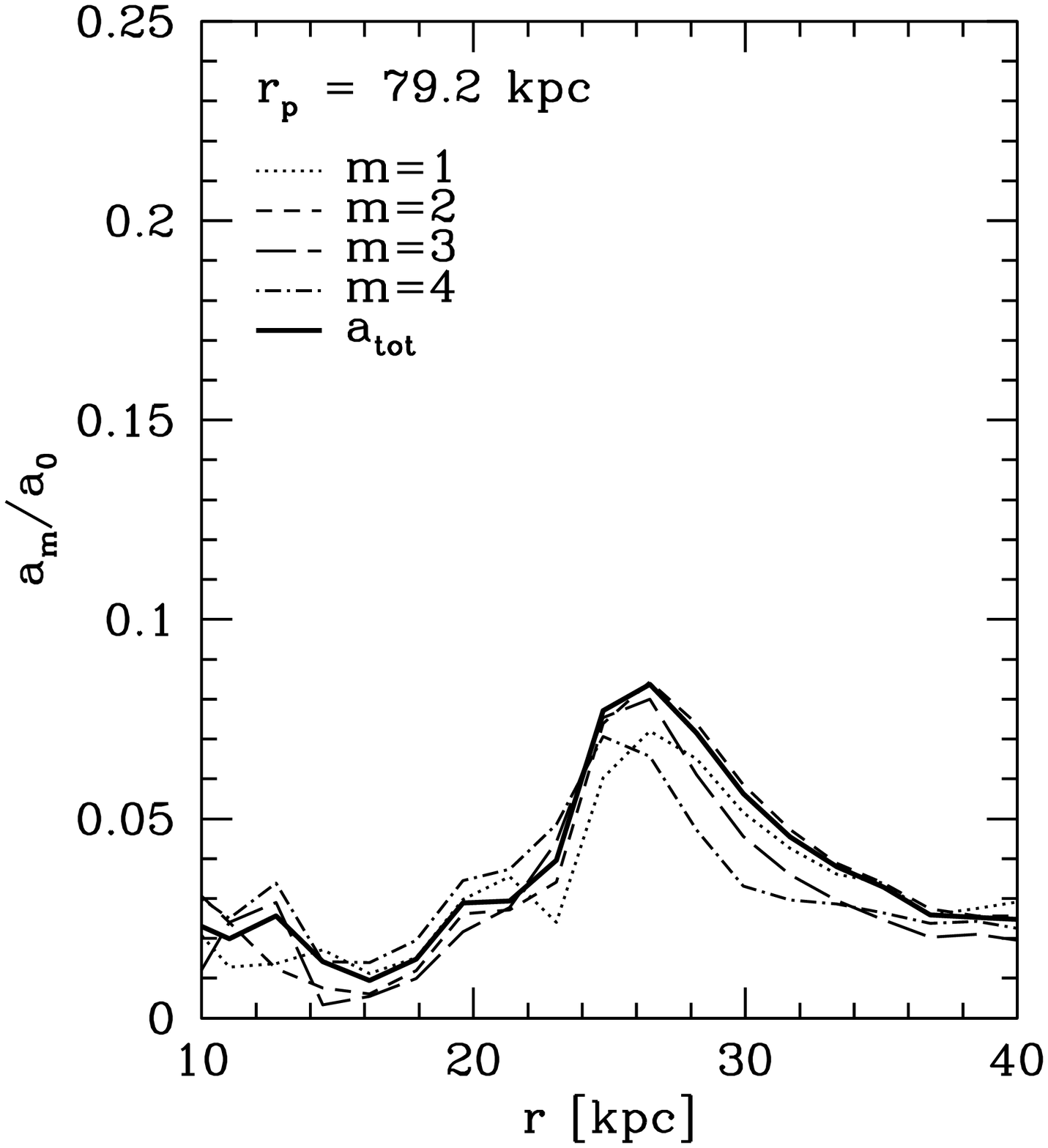}}
\subfigure[]{
\includegraphics[width=.23\textwidth]{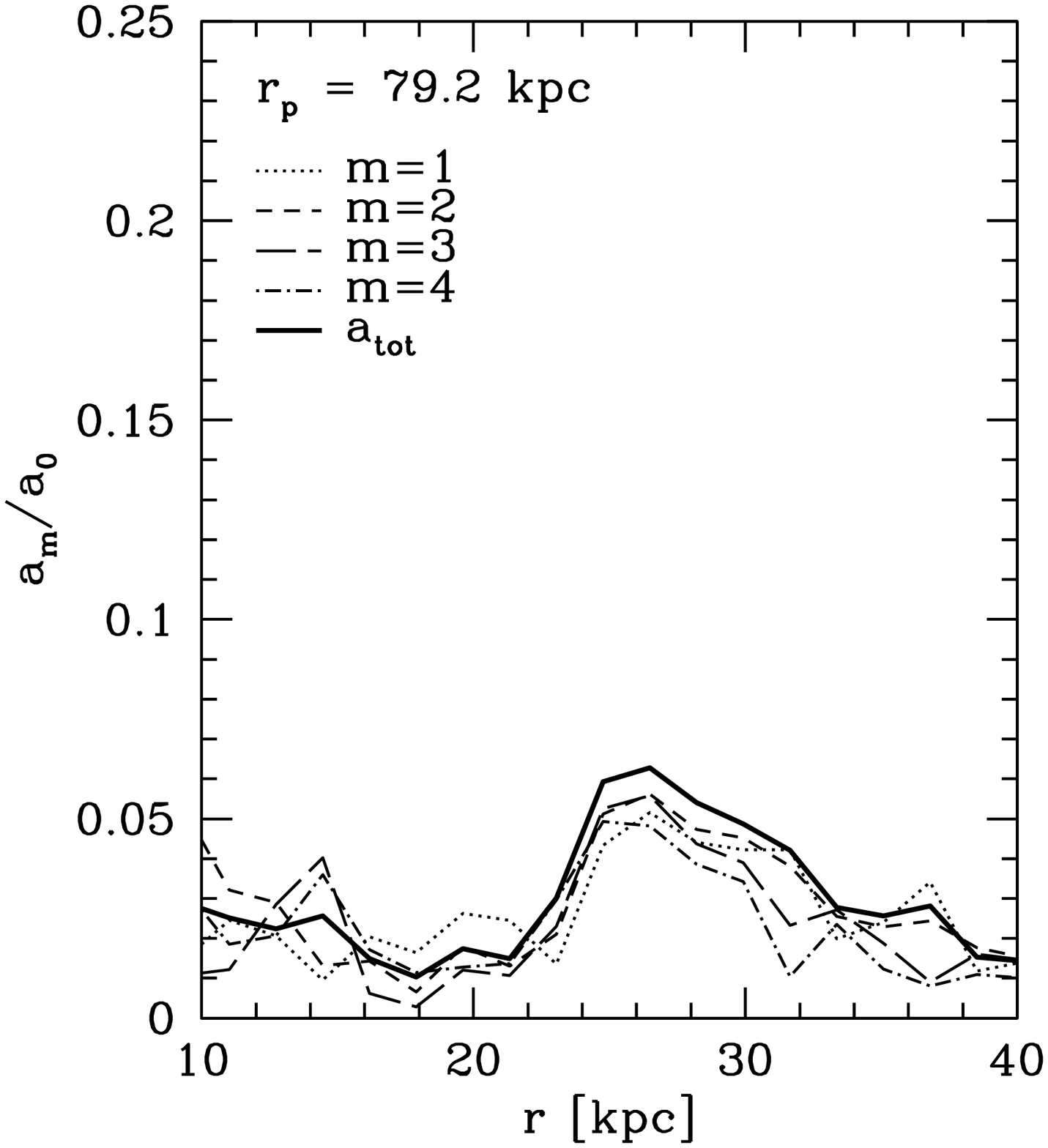}}
\subfigure[]{
\includegraphics[width=.23\textwidth]{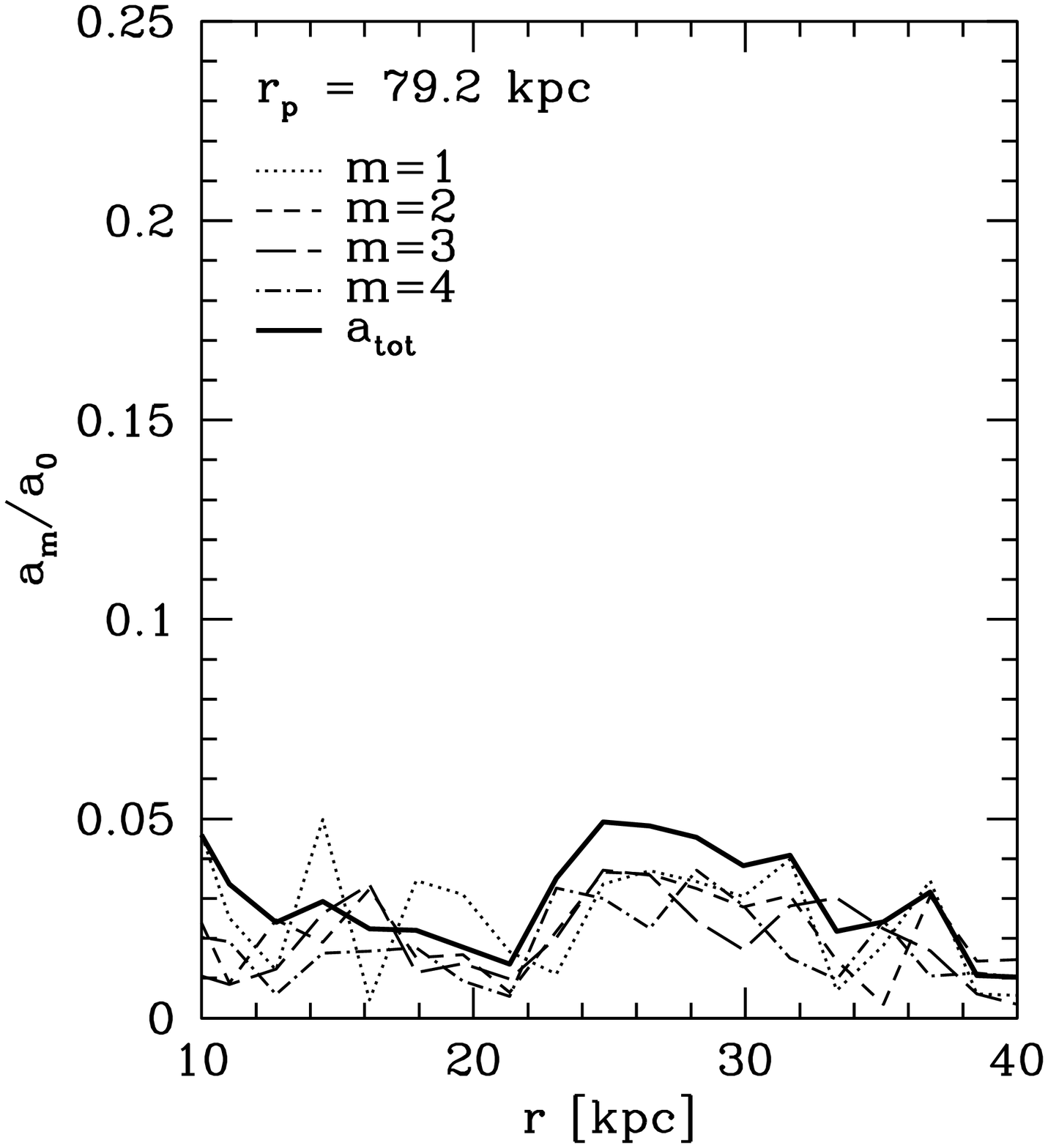}}
\subfigure[]{
\includegraphics[width=.23\textwidth]{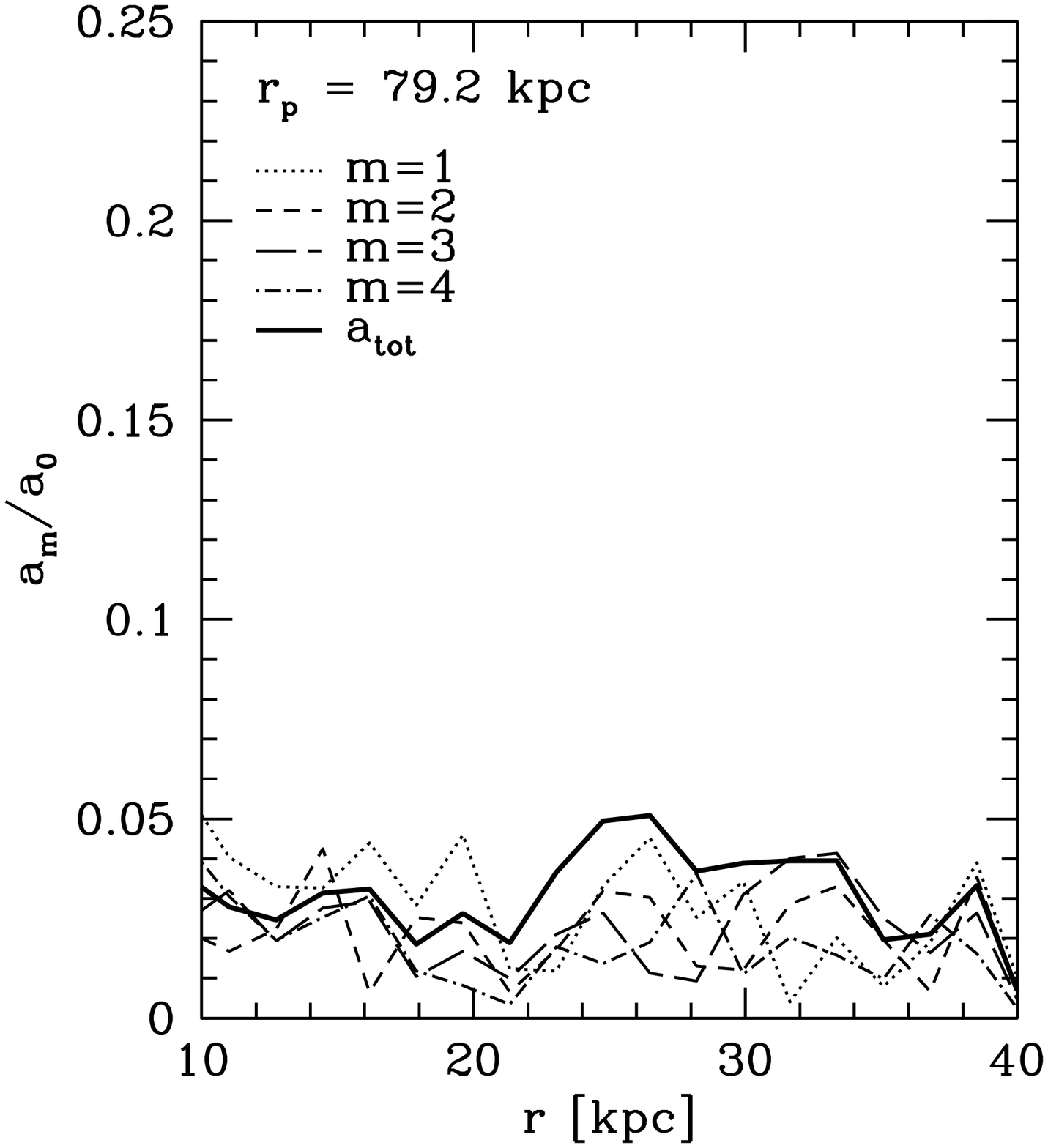}}
\subfigure[]{
\includegraphics[width=.23\textwidth]{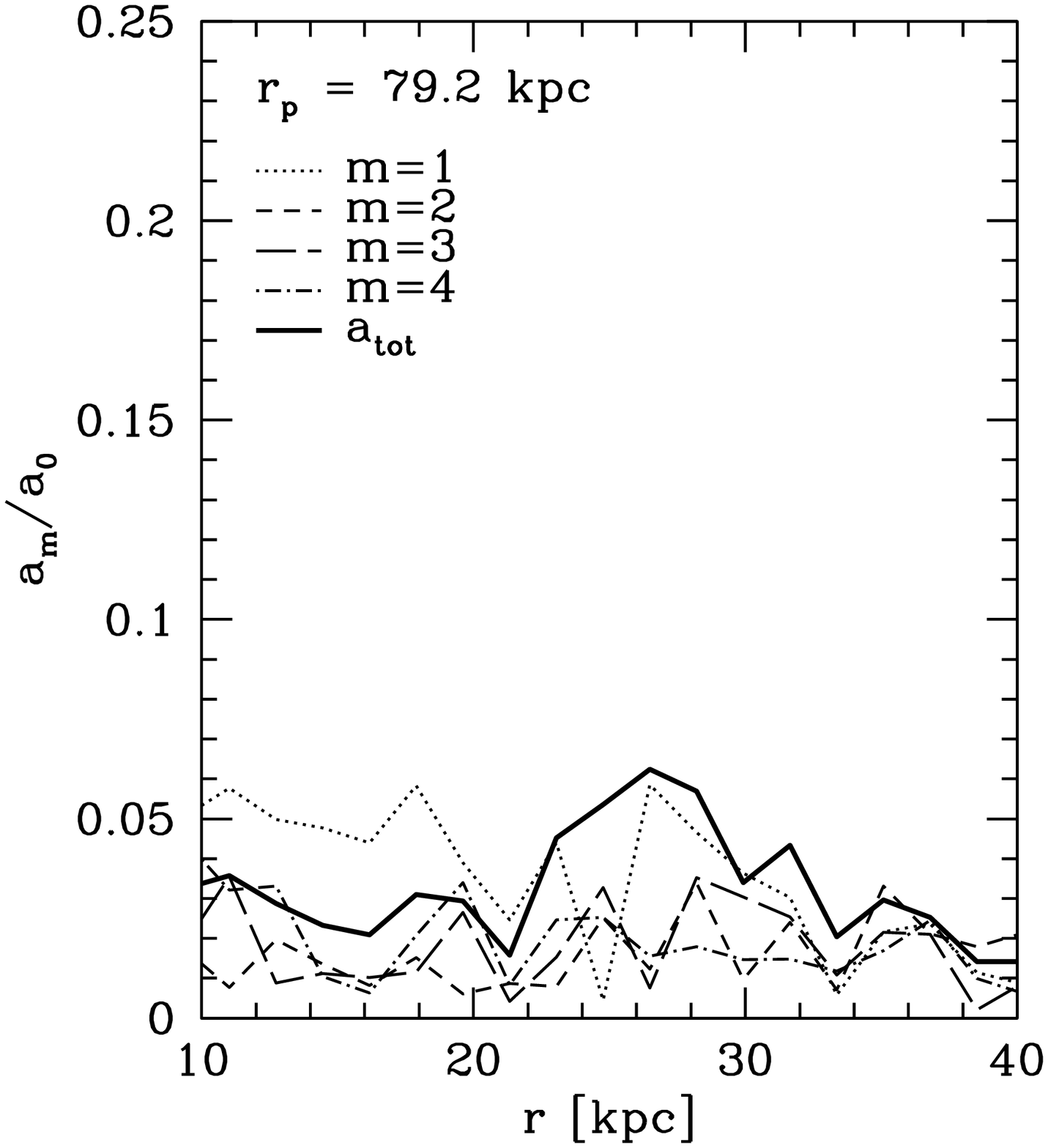}}
\subfigure[]{
\includegraphics[width=.23\textwidth]{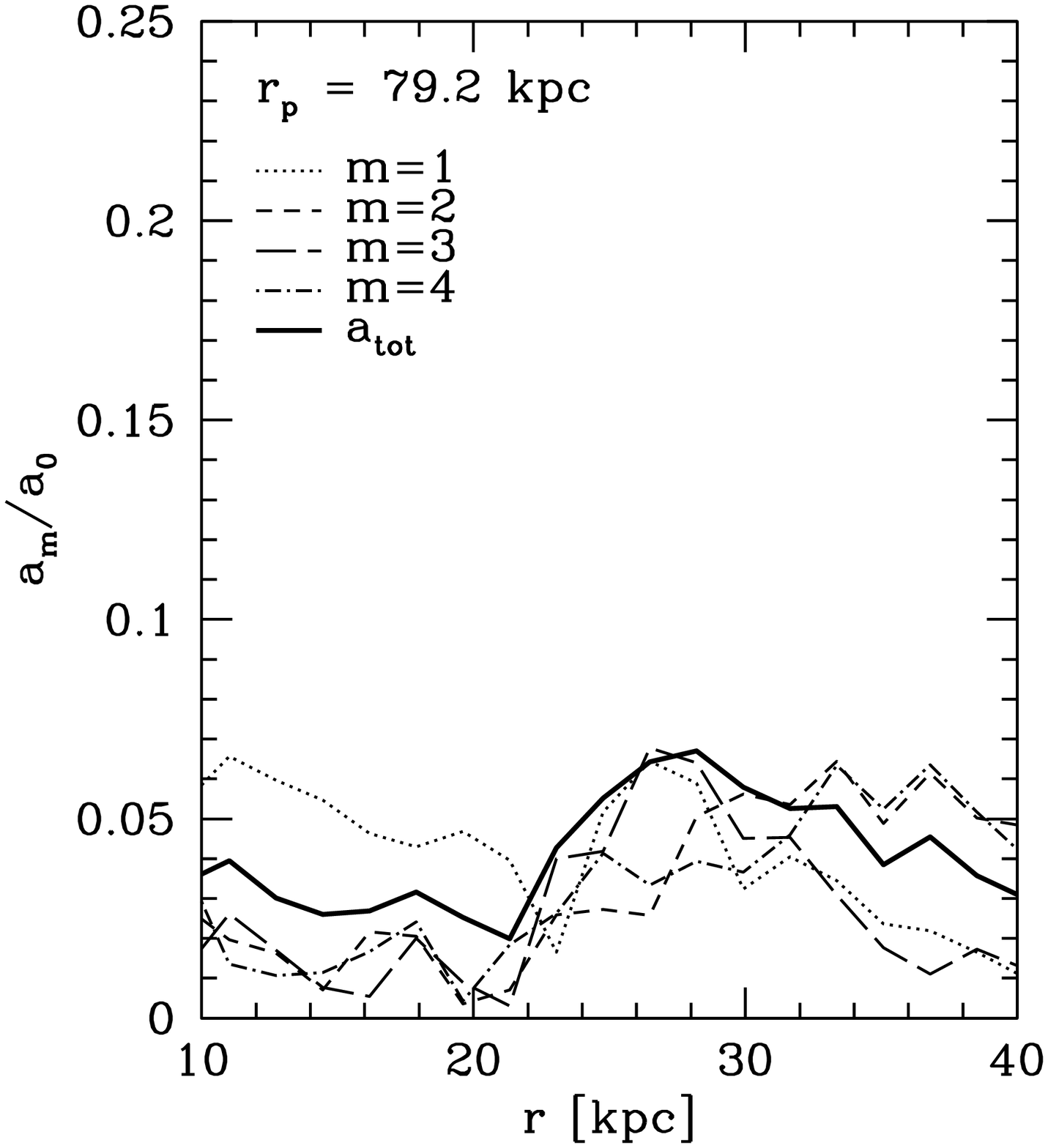}}
\end{center}
\caption{ Snapshots of 1:100 perturber $r_{\rm peri} = 20$ kpc for different
inclinations. For each inclination, we have taken the same snapshot, which
corresponds to (f) of Figure \ref{f:dens100_20kpc}. The density response peaks around $r = r_{\rm peri}$ in each of these
snapshots, in line with the expectations from our energy analysis.}
\label{f:dens100_inclination}
\end{figure*}

We now use this synthetic measure (eq.[\ref{eq:atot}]) to show that this localized response around $\rperi$ continues to hold as we vary the orbit with different inclinations and different angles.   In Figure \ref{f:dens100_inclination}, we show the effects of different inclination on the Fourier power of the resulting density structure, while holding angle fixed.  Comparing Figure \ref{f:dens100_inclination} with \ref{f:Em100_inclination}, we find that again that the rapid increase in the $m=0$ power is a good discriminant for $\rperi$.  The $m\neq 0$ modes also show good discriminating power though not as the $m=0$ mode.  However, what is really encouraging is that, while the $m=1$ mode varies signficantly with inclination in Figure \ref{f:Em100_inclination}, its effect on the density response is much milder.  From looking at these plots, it is clear that the $m=0$ modes has the most discriminating power, but $a_{\rm tot}$ appears to be adequate for determining \rperi.  
 
\section{Scaling Relations: Inferring Satellite Mass from Fourier Amplitudes}\label{sec:scaling}

Our discussion in \S~\ref{sec:influence} shows while the overall dynamical response of the disc depends on the orbital parameters, i.e., $\rperi$, inclination and angle, the density response of the disc is generally localized around $\rperi$ and \rperi\ + 10 kpc.  This localized response, especially for the $m=0$ mode allows us to constrain \rperi\ of the orbit.  We now discuss the effect of the subhalo's mass on the disc response with an aim toward developing scaling relations between the response and the perturber's mass.  

\begin{figure}
  \includegraphics[width=0.45\textwidth]{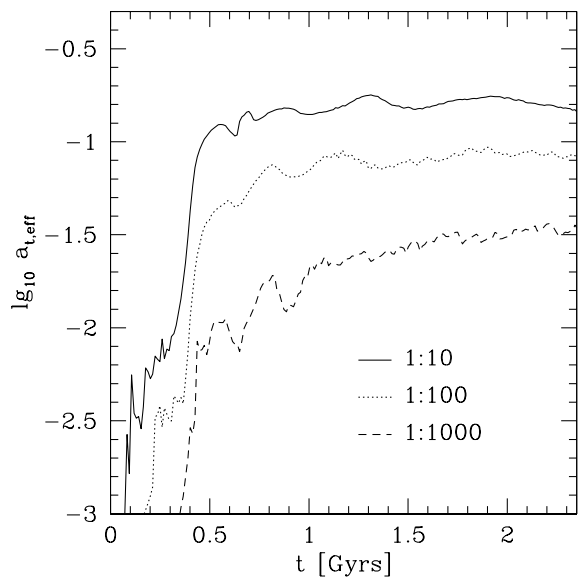}
  \caption{Plot of the synthetic Fourier amplitude of the density response as a function of time.  Note that after the initial interaction, the amplitude stay relatively constant in the absence of gaseous dissipation.  Whereas this is expected for the modal energies, it is surprising that this is preserved in the density response as well.  It points out that the synthetic amplitude is a good proxy for measuring the global properties of the response in a robust manner.}
\label{f:atot}
\end{figure}

We begin by defining an appropriate global measure of the response of the disc in the much the same spirit as our earlier definition of $a_{\rm tot}$ (eq.[\ref{eq:atot}]).  As we have seen already that the disc response is localized, we define the effective amplitude of the disc as
\begin{equation} \label{eq:effective amplitude}
a_{\rm m, eff} = (\Delta r)^{-1}\int_{\rperi}^{\rperi+\Delta r} |a_m(r)| dr,
\end{equation}
where $\Delta r = 10$ kpc, motivated by the results of the previous section.  We then define the total effective amplitude as 
\begin{equation}
a_{\rm t, eff} = \sqrt{\frac 1 4 \sum_{m=1}^4 |a_{\rm m,eff}}|^2.
 \end{equation}
A major advantage of defining such global quantities is that like the modal energies, their overall amplitudes settles into a relatively constant range after the initial excitation, in the absence of dissipation.  We show this in Figure \ref{f:atot}, for DM perturbers with mass ratios of 1:10, 1:100, and 1:1000.  Note that $a_{\rm t,eff}$ settles into a constant range that depends on the mass ratio $\approx 1$ Gyr after their interaction,  which is roughly one rotation period of the outer disc. 

\begin{figure}
  \includegraphics[width=0.5\textwidth]{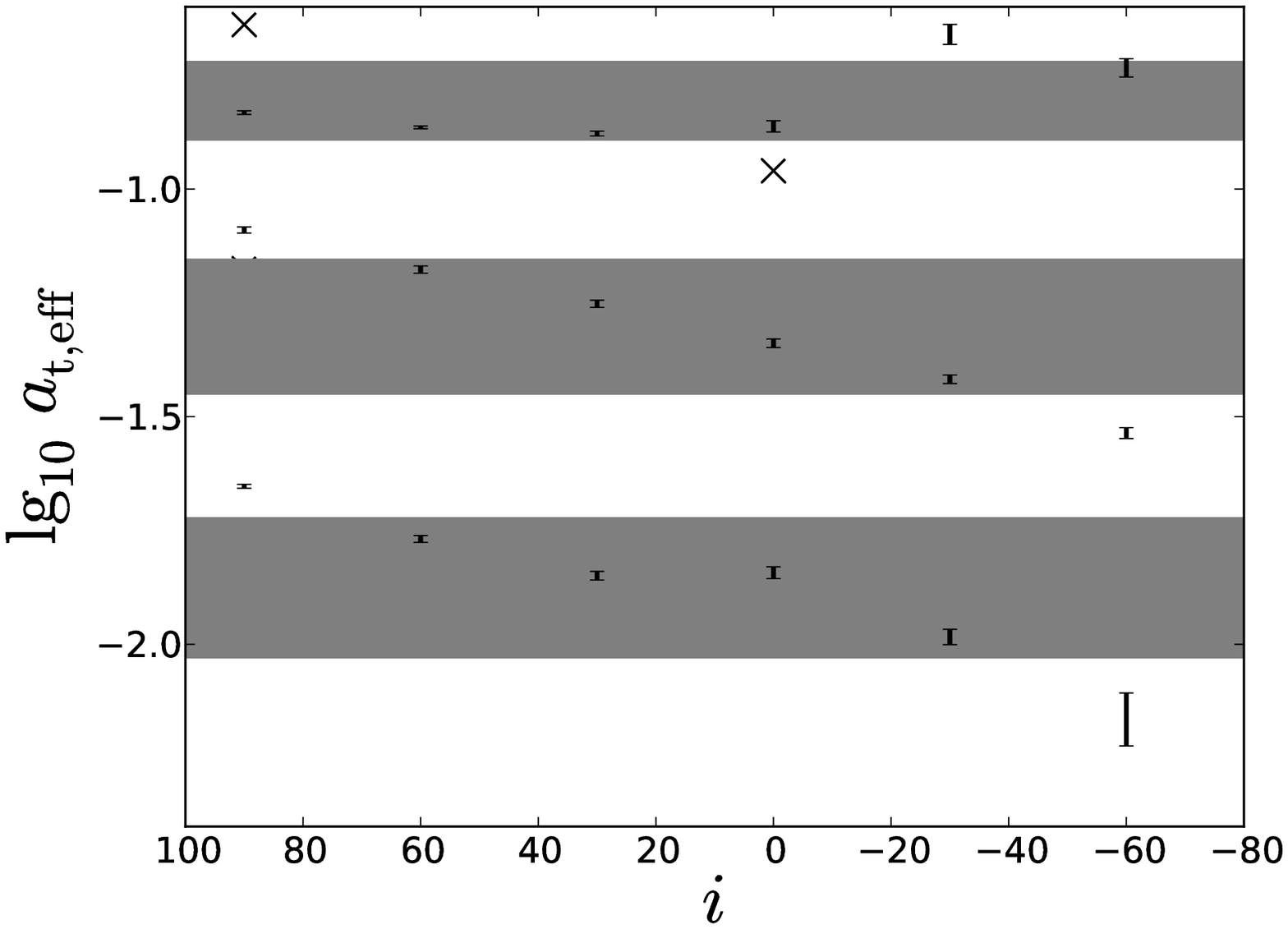}
  \caption{Late time amplitude as a function of the inclination for 1:10 (top set of points), 1:100 (middle set of points), and 1:1000 (bottom set of points).  The error bar represent the 1-$\sigma$ spread in the synthetic amplitude as the angle is varied and the banded regions represents the 1-$\sigma$ spread in the range of synthetic amplitude as inclination is varied.}
\label{f:atot_inc}
\end{figure}

\begin{figure}
  \includegraphics[width=0.5\textwidth]{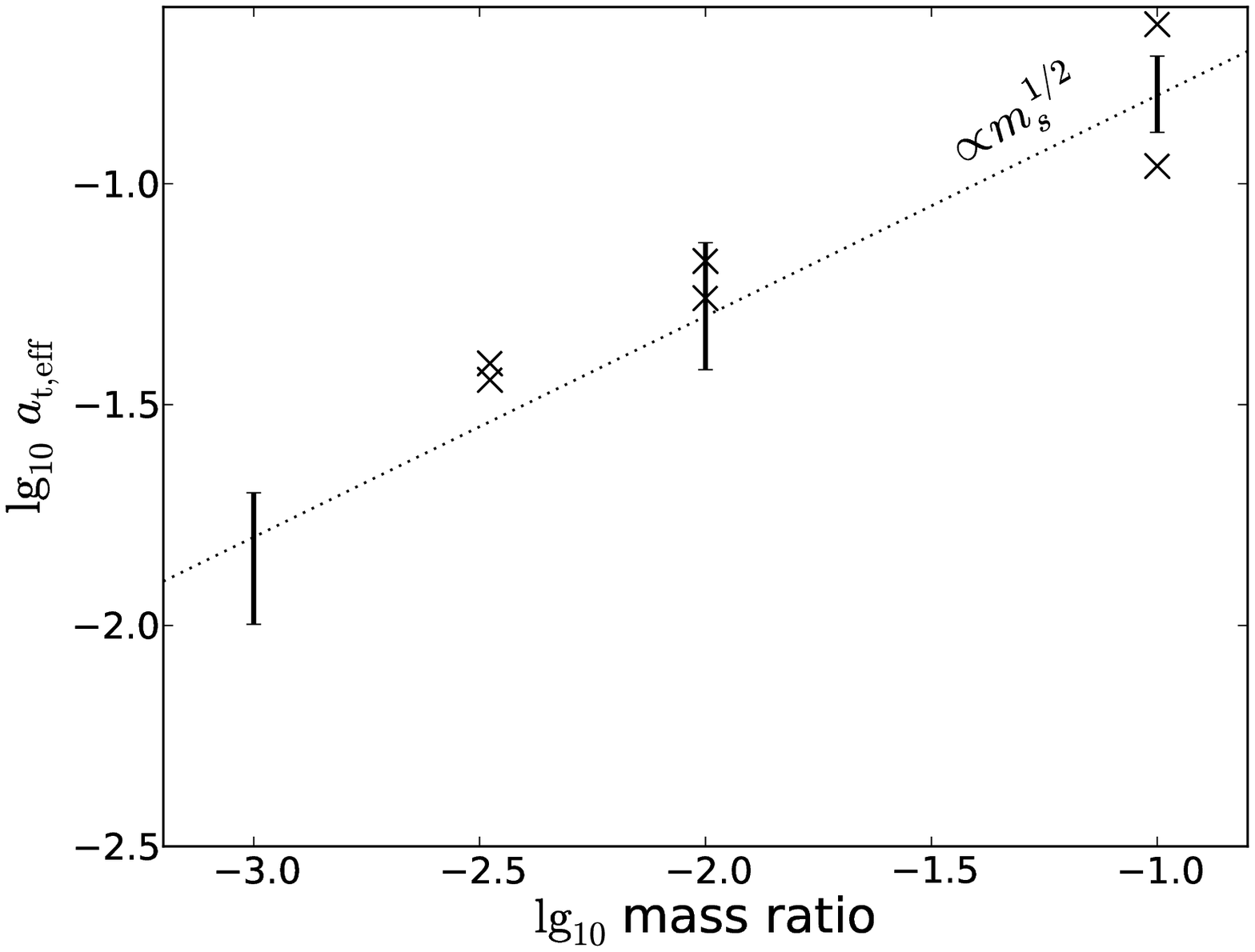}
  \caption{The central and the ranges of $a_{\rm t,eff}$ as a function of mass ratio, varying from 1:1000 (left point) to 1:10 (right points).  The x's show the results from the full SPH simulations described in \S \ref{sec:sph}. The dotted line shows a relation that scales like $m_s^{1/2}$.  }
\label{f:atot_mass}
\end{figure}

Motivated by this result, we average $a_{\rm t,eff}$ after $\sim$ one rotation period for different inclinations and angles in Figure \ref{f:atot_inc}.  The error bars define the variation in $a_{\rm t,eff}$ for different angles.  The bands represent the variation of $a_{\rm t,eff}$ for each mass ratio as we vary the inclination angle from 90 to -90.  The variation in the inclinations is  $\approx 2$.  However, the average amplitude over which $a_{\rm t,eff}$ varies depends strongly   
on the mass ratio of the satellite.  We show this mass ratio dependence more clearly in Figure \ref{f:atot_mass} where we collate the results of Figure \ref{f:atot_inc} and plot the overall disc response as a function of $m_{\rm s}$.  The error bars represent the variation over inclination and angle (same as the banded structures in Fig \ref{f:atot_inc}). In addition we plot the following fitting formula (dotted line)
\begin{equation}\label{eq:fit}
 a_{\rm t,eff} = 0.5 \sqrt{\frac {m_s} M}.
\end{equation}
Equation (\ref{eq:fit}) provides a reasonable fit to the numerical results as shown by the dashed line in Figure \ref{f:atot_mass}.  It suggests that the energy of the modes, $E \propto a^2 \propto M_{\rm s}$.  This is different from the expectations of the impulse approximation which instead suggest that the $E \propto M_{\rm s}^2$ \citep{White00}.  The fact our results differ from the impulse approximation means that the impulse approximation is not a good approximation for the interaction of subhaloes with extended HI discs, a point already discussed by CB09, CB10, and CBCB.   

\section{Comparison to SPH Simulations}\label{sec:sph}

To confirm our analysis, we compare our results from the modal analysis above to full-scale numerical galactic simulations.  We use the smooth particle hydrodynamic code, Gadget-2 \citep{Springel05} using the methodology presented in CB09.    GADGET-2 uses an N-body method to follow the evolution of the collionsionless components, and SPH to follow the gaseous component.  The simulations of disc galaxies tidally interacting with dark matter mini-halos reported here (unless otherwise noted) have gravitational softening lengths of $100~\rm pc$ for the gas and stars,
and $200~\rm pc$ for the halo.  The number of gas, stellar, and halo particles in the primary galaxy are $4 \times 10^{5}$, $4 \times 10^{5}$ and $1.2 \times 10^{6}$ respectively for our fiducial case. We refer the interested reader to CB09 for additional details on the simulation methodology.

We focus here on a few representative cases that span a reasonable range in parameter space. We list in Table 1 the parameters for the simulations that we use for this comparison. They vary in mass ratio between 1:10 and 1:300 and have coplanar and polar inclinations.  

\begin{figure*}
  \includegraphics[width=0.9\textwidth]{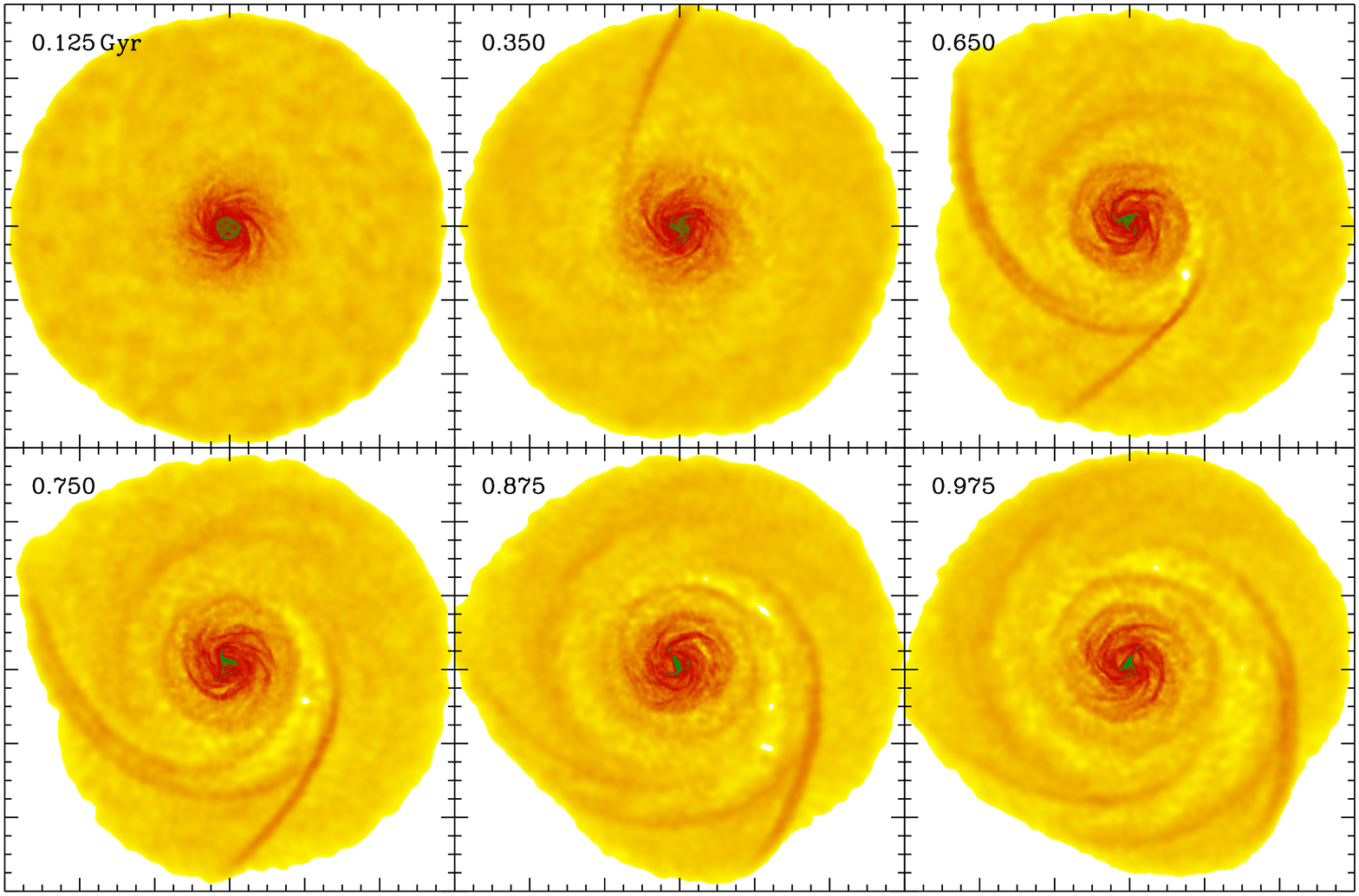}
  \caption{{The time evolution of the gaseous density images of a Milky Way like galaxy interacting with a 1:100 dark sub-halo with a pericentric approach distance of 33 kpc. The units of time, marked in the figure, are in Gyr.  The box extends from -60 kpc to 60 kpc. }}
\label{f:SPH-1:100}
\end{figure*}

As an example, Figure \ref{f:SPH-1:100} shows the density response for a few time snapshots of the interaction for the 1:100 coplanar example.  The general response of the disc in the full simulation is visually similar our modal calculations of Figure \ref{fig:nbody_vs_analytic} especially that of $t = 0.650$ Gyrs.  This is somewhat unsurprising as the dominant physics in both calculation is the gravity of the primary halo and the subhalo.  However, it does highlight that much of the physics that has been ignored in the simplified calculation, i.e., a live dark matter halo, tidal stripping of the subhalo, gas dissipation, star formation, and self-gravity, 
does lead to some difference in the appearance of the disturbances in the HI disc. 
A main physical effect that has not been taken into account in these calculations is gas dissipation.  Neglecting this effect allows for simplicity of calculation as the modal energy (and the Fourier amplitudes once generated) are constant in the absence of dissipation.  In the presence of gas dissipation, the Fourier amplitudes decrease as a function of time as disturbances in the gas disc damp out on the order of a dynamical time.  We do not model this effect here, which would be needed to identify the current location of a satellite from observed disturbances in the HI disc as done earlier in the SPH calculations by CBCB. 

Motivated by the visual agreement of the SPH and modal calculation, we focus on comparing $a_{\rm t,eff}$ for the SPH simulation and the modal calculation.  Here the agreement is striking.  In the final column of Table 1, we list the $a_{\rm t,eff}$ for the various SPH simulations that we have performed.  We also $a_{\rm t,eff}$ for the SPH simulations as x's in Figure \ref{f:atot_inc} and \ref{f:atot_mass}.  The SPH simulation and the modal calculation show striking agreement especially for the 1:100 case.  We have also computed the 1:300 case and it too shows excellent agreement between the modal calculation and the SPH calculation.  The 1:10 case show some mild disagreement between the SPH simulation and the simplified modal calculation.  This is not unexpected as linear theory should start to break down for 
sufficiently massive satellites (mass ratios $\gtrsim$ 1:10).  The overlaid points from the SPH simulation in Figure \ref{f:atot_mass} gives additional support to the $a_{\rm t,eff} \propto m_s^{1/2}$ scaling that we initially found from the modal calculation.

\begin{table}
\label{table:sim}
 \centering
 \begin{minipage}{90mm}
  \caption{Parameters of SPH simulations and $a_{\rm e,tot}$}
  \begin{tabular}{rrrrrrrr}
  \hline
   mass ratio &  $\rperi$ [kpc] & orientation & $a_{\rm e,tot}$\\
  \hline
      1:10    &   30  & coplanar & 0.22\\
                   &         & polar       & 0.11\\
      1:100   &   32  & coplanar & 0.061\\
                   &          & polar        &   0.055 \\
      1:300   &   34  & coplanar & 0.039\\
                   &          & polar        &  0.036    \\
\hline
\end{tabular}
\end{minipage}
\end{table}

\section{Conclusion and Discussion}\label{sec:conclusions}

Subhalo interactions with the HI discs of galaxies should leave measurable effects on the gas long after the subhalo has passed by as demonstrated by CB09, CB10, and CBCB.  In this paper, we look carefully at the dynamics of these interactions in the context of a simplified epicyclic approximation.  

$\bullet$ By performing a modal analysis on the equations of motion, we show that we can reproduce the orbital dynamics
and distribution of particles of a test-particle calculation for identical conditions.  While this may not seem surprising given the mass range of the perturbers that we are interested in ($\lesssim 0.1$), we show that performing such an analysis, we can understand crucial aspects of the nature of such interations.  In particular, we show that the modal energy shows sharp variations at $\sim r_{\rm peri}$.    

$\bullet$ 
By developing a sense of what to look for from the modal analysis, we then showed that such variations also show up in the Fourier modes of the $\it{perturbed~density}$.  Namely, we find sudden increases in the Fourier modes of the density response at $\sim r_{\rm peri}$.  Such sharp variations at large radii are a distinct signature of tidal interaction and would argue against a secular origin.  

$\bullet$ We also showed that this simplified analysis agrees with test particle calculations as well as more extensive SPH simulations.  This demonstrates that the physics ignored in the simplified approach (a live dark matter halo, tidal stripping of the subhalo, gas dissipation effects, self gravity and star formation, and dynamical friction) do not significantly alter the 
production of the 
disturbances in the HI discs.  

$\bullet$ One of our main results here that can be readily employed by observers is that the effective Fourier amplitdues of the resultant surface density scales as the square root of the mass of the satellite, i.e., $a_{\rm t,eff} \propto m_{\rm s}^{1/2}$.  This suggests that the energy of the modes is $\propto m_{\rm s}$.  This scaling relation can be used to analyze observed HI maps and detemine the mass of the perturbing satellite without having to take recourse to full numerical simulation.


In future work, we will apply these scaling relations to analyze results from cosmological simulations to determine how the dark sub-halos will impact the disc.  To do this, we will generalize our method to incorporate multiple perturbers.
The effect of multiple perturbers is not yet known, but it may be the case that multiple perturbers do not significantly alter our results.  Namely, if the timescale between encounters is significantly longer than a dynamical time, the disturbances in the gas disc from the first encounter will be damped out.   This $\it{short-term~memory}$ of the gaseous component allows us to more cleanly disentangle the effect of the last perturber from the ones that impacted the disc previously.  Simulations predict impacts with $\sim 1:100$ mass ratio perturbers occur only once every $\sim \rm Gyr$ \citep{Kazantzidis+08}, which bodes well for this possibility.  

\section*{Acknowledgments}
We thank Leo Blitz and Eric Gawiser for helpful discussions.  P.C. is supported by the Canadian Institute for Theoretical Astrophysics.  

\bibliographystyle{mn2e} 
\bibliography{subhaloes}

\end{document}